\documentclass[epj]{webofc}
\usepackage[varg]{txfonts}   % Web of Conferences font
\wocname{EPJ Web of Conferences}
\woctitle{New Frontiers in Physics 2014}
\newcommand{\be}{\begin{equation}}
\newcommand{\ee}{\end{equation}}
\newcommand{\bea}{\begin{eqnarray}}
\newcommand{\eea}{\end{eqnarray}}

\newcommand{\6}{\partial }

\newcommand{\lSS}{\bar\lambda} %{\lambda_{\rm SS}}
\newcommand{\MKK}{M_{\rm KK}}

\newcommand{\uKK}{u_{\rm KK}}
\newcommand{\gYM}{g_{\rm YM}}

\newcommand{\Tr}{{\rm Tr}\,}

\begin{document}
\selectlanguage{english}
\title{The Witten-Sakai-Sugimoto model:\\ A brief review and some recent results}
%Holographic glueball decay and other recent results}
%\subtitle{Holographic glueball decay and other recent results}
%
% subtitle is optional

\author{Anton Rebhan\inst{1}\fnsep\thanks{\email{rebhana@tph.tuwien.ac.at}}
}

\institute{Institute for Theoretical Physics, TU Wien\\
Wiedner Hauptstr. 8--10, A-1040 Vienna, Austria}

\abstract{%
  A brief review of the Witten-Sakai-Sugimoto model is given, which is a top-down holographic model of low-energy QCD with chiral quarks derived
  from type-IIA superstring theory. The main predictions of the model, in particular concerning meson spectra, the gluon condensate,
  the QCD string tension, the mass of the $\eta'$ and of baryons are discussed and compared quantitatively with available experimental and/or lattice results.  Then some recent results of potential interest to the physics program at the future FAIR facility
  are presented: The spectrum of glueballs and their decay rates into pions, and the phase diagram of QCD at finite temperature, density, and magnetic field strength.
}
\maketitle
\section{Introduction}
\label{intro}

Quantum chromodynamics (QCD) is the well-established theory of the strong nuclear forces. Thanks to asymptotic
freedom, it is amenable to perturbative methods at high energy. While the coupling never becomes
very small at any energy scale of interest, perturbative QCD is a most successful framework for collider physics,
and its predictions beyond the simple parton model have been tested extensively.
Nonperturbative aspects of QCD, such as the spectrum of hadrons and the transition to a quark-gluon plasma phase,
can be studied from first principles through lattice QCD. Indeed, tremendous progress has been made thanks
to the availability of powerful computers and also thanks to methodological progress. However, two areas of strong-coupling
physics, where lattice QCD is still far from giving reliable answers are finite baryon density and
real-time phenomena such as decay rates (although progress is being made on both frontiers).
These insufficiently charted areas will be at the focus of a new generation of experiments at the FAIR
facility, where the PANDA experiment at FAIR will study hadron reactions \cite{Lutz:2009ff,Wiedner:2011mf}, 
and the CBM experiment will
aim at exploring baryonic matter at high density \cite{Friman:2011zz}.

A novel approach to strongly coupled gauge theories has been developed over the last one and a half decades
in the form of gauge/gravity duality \cite{Aharony:1999ti}, also known as ``holography''. While first-principles approaches with
a well-defined string-theoretic foundation allow one to study only gauge theories which are in important
ways different from real QCD, this approach offers fresh insights in the richness of phenomena of gauge theories
at strong coupling. Based on this progress, so-called bottom-up models have been developed which do not
care so much about a string-theoretic justification, but provide phenomenological models that incorporate
the key ingredients of holographic gauge theories, and many interesting studies have been carried out using these.

There is however one top-down holographic model, which goes a long way towards QCD starting from a
fundamental string-theoretic basis: Witten's model
of low-energy QCD through type-IIA supergravity, proposed already in 1998 \cite{Witten:1998zw}, which has almost no free parameters. 
Here one sets up a holographic dual to a supersymmetric
gauge theory in 4+1 dimensions and performs the analogue of high-temperature dimensional
reduction, but with respect to the superfluous spatial dimensions, such that at low energy one ends
up with nonsupersymmetric pure-glue QCD at large number of colors and large 't Hooft coupling in 3+1 dimensions.

In 2004, Sakai and Sugimoto \cite{Sakai:2004cn,Sakai:2005yt} have extended Witten's model by flavor D8-branes in a setup that
provides chiral quarks and anti-quarks. In the model of Sakai and Sugimoto chiral symmetry breaking
as well as its restoration at high temperatures has a beautiful geometric realization. Its low-energy
effective action contains a chiral Lagrangian for massless Nambu-Goldstone bosons and massive vector
and axial vector mesons, complete with the correct Wess-Zumino-Witten term and also a Skyrme term.
The only free parameters in the model are a (Kaluza-Klein) mass parameter
$M_{\rm KK}$ and the 't Hooft coupling $\lambda$ at this scale, making it
very predictive. Unfortunately, hardly anything is known about the corrections that are needed
to approach real QCD at finite $N_c$ and $\lambda$ and that are in principle
determined by type-IIA string theory. In the case of the original AdS/CFT correspondence for 
maximally supersymmetric Yang-Mills theory based on type-IIB string theory, 
higher-derivative string-theoretic corrections
%in inverse powers of $N_c$ and $\lambda$ 
have been determined for entropy and shear viscosity \cite{Myers:2008yi}.
Reassuringly, they come with the expected sign to connect to perturbative results.\footnote{In the case of the entropy of a supersymmetric plasma,
it is has even been found that the next-to-leading results at weak and at strong coupling
can be interpolated uniquely with a singularity-free Pad\'e approximant \cite{Blaizot:2006tk}.} For $N_c=3$ and $\alpha_s=0.5$, their magnitude is
5\% and 37\% for  entropy and shear viscosity, respectively,
which (optimistically) gives an idea of the quantitative predictiveness to expect in top-down holographic
calculations. 
Indeed, as will be shown below, the Sakai-Sugimoto model does not only reproduce
many qualitative features of QCD but is often quantitatively close. This makes it interesting to
apply it to new questions that are presently out of reach of a first-principles approach such as lattice QCD.
After a brief review of the Witten-Sakai-Sugimoto model and how it compares quantitatively with
real QCD, we will discuss some new results
obtained for the spectrum and decay rates of glueballs, and for the phase diagram of QCD in the presence
of strong magnetic fields (for the latter see also the recent reviews \cite{Kharzeev:2013jha}).

\section{Brief review of the Witten-Sakai-Sugimoto model}
\label{secWSS}

\subsection{Large $N_c$ gauge theories and the AdS$_5$/CFT$_4$ correspondence}

The dominant Feynman diagrams of nonabelian gauge theories in the limit of large number of colors
but fixed 't Hooft coupling $\lambda=g^2 N_c$ are the planar diagrams, which can be drawn on a plane (or, equivalently, a sphere) without
crossing of lines when the color flow
is represented by one line per fundamental or antifundamental index (so that gluons are represented
by double lines). Nonplanar diagrams are suppressed by a factor $(1/N_c)^{\,\chi}$, where $\chi$ is
the Euler number of the surface needed to draw the Feynman diagram.\footnote{A nice illustration
is provided by Fig.~2 of Ref.~\cite{Peeters:2007ab}.} This is very similar to string perturbation
theory, where one has to sum over Riemann surfaces weighted by $g_s^{\,\chi}$ with $g_s$ the string coupling constant, 
leading 't Hooft to the
remarkable speculation that the large-$N_c$ limit of nonabelian gauge theories may be essentially
a string theory \cite{'tHooft:1973jz}. 

A quarter of a century later, this idea became
finally concrete in the form of the AdS/CFT correspondence.
The study of nonperturbative objects in string theory, so-called D-branes, led Maldacena \cite{Maldacena:1997re} to
the conjecture that string theory in the near-horizon limit of $N_c$ coincident D3-branes, which is governed by
a curved ten-dimensional space of the form AdS$_5\times S^5$, is completely equivalent to the low-energy
effective theory of these branes in flat Minkowski space, which is the maximally supersymmetric
conformal SU($N_c$) gauge theory. Superstring theory, which necessarily involves gravity, in
the background of 5-dimensional anti-de Sitter space times a 5-sphere of equal curvature radius
is thus proclaimed to give an alternative description of a nonabelian quantum gauge theory 
in 4 spacetime dimensions without gravity. The latter can be viewed to live on the 4-dimensional
(conformal) boundary of anti-de Sitter space, which thus also realizes the holographic principle
anticipated for quantum gravity by 't Hooft and Susskind \cite{Susskind:1994vu}.

Spatially three-dimensional branes are nonperturbative objects in type-IIB superstring theory as well
as in its low-energy limit, type-IIB supergravity. The latter is a justified approximation when
the string length scale $\ell_s=\sqrt{\alpha'}$
is small compared to spacetime curvature radius\footnote{Not to be confused with
the Ricci scalar, which will be written as $R^{(d)}$ for a $d$-dimensional space.} $R$ and when the string coupling $g_s$ is
small. Because the weak-coupling and strong-coupling description of D3 branes give, respectively,
rise to the two relations
\begin{equation}
 g^2=4\pi g_s,\qquad 4\pi g_s N_c=\frac{R^4}{\ell_s^4},
\end{equation}
a useful gauge/gravity duality emerges in the limit $g^2\to 0$ and $\lambda\gg 1$, implying $N_c\to\infty$.
Useful, because it allows one to substitute extremely difficult
calculations in the strongly coupled quantum gauge theory by
calculations in weakly-coupled, classical supergravity. (It should be noted, however,
that the AdS/CFT correspondence is at its roots a duality between two quantum theories -- superstring theory
in a certain background on the one hand, and a superconformal field theory on the other.)

The gauge theory in question is however a (maximally) supersymmetric and conformal quantum field theory
which is very different from QCD, which is neither supersymmetric nor conformal, but has a running coupling constant
and confinement. However, for deconfined QCD at high temperature,
the AdS/CFT correspondence may actually be useful. The super-Yang-Mills theory can be considered at finite temperature, where
the geometry on the supergravity side becomes one that involves 
a black hole (strictly speaking a black brane) in
anti-de Sitter space. Finite temperature breaks supersymmetry because bosons and fermions
have different distribution functions, whereas conformal symmetry may actually be a reasonable approximation
sufficiently above the deconfinement temperature. Indeed, the quark-gluon plasma created in
heavy-ion collisions seems to behave in many respects like a strongly coupled quantum fluid. In particular
hydrodynamic transport coefficients such as shear viscosity turn out to be better described by the
famous result \cite{Policastro:2001yc,Kovtun:2004de} 
$\eta/s=\hbar/4\pi$ obtained through the AdS/CFT correspondence than by perturbative QCD, where
$\eta/s$ is parametrically large.
Perturbative QCD is certainly relevant for high-energy processes, and it seems most promising to combine
perturbative and strong-coupling gauge/gravity approaches when both have
a role to play such as in high-energy jets that are quenched by a strongly coupled medium \cite{Casalderrey-Solana:2014wca,Iancu:2014ava}.

\subsection{Towards a gravity dual of QCD: the Witten model}

At low temperature, with and without quark chemical potential, superconformal Yang-Mills theory is
however certainly very different from real QCD.
An intriguing route towards a gravity dual of QCD has been proposed by Witten in Ref.~\cite{Witten:1998zw}:
There are two more fundamental instances of AdS/CFT dualities, arising from the so-called
M2- and M5-branes of M-theory, whose low-energy limit is the (unique) 11-dimensional supergravity,
already discussed by Maldacena in Ref.~\cite{Maldacena:1997re}.
The near-horizon geometries of these branes are AdS$_4\times S^7$ and AdS$_7\times S^4$, respectively,
which were found as solutions of 11-dimensional supergravity already in 1980
by Freund and Rubin \cite{Freund:1980xh} when they were searching for Kaluza-Klein reductions to 4 spacetime 
directions.\footnote{These solutions are stabilized by
having constant flux from the antisymmetric tensor fields of 11-dimensional supergravity on the compact spheres, 
for which the M-branes provide sources.
Because the anti-de Sitter space and the sphere have comparable curvature, these
solutions are not viable as Kaluza-Klein theories for the real world, but they paved the way
to more elaborate flux compactifactions in string theory.}
The dual of M-theory on AdS$_7\times S^4$ is the 6-dimensional superconformal
low-energy theory of M5 branes. 
M-theory compactified on a supersymmetry-preserving circle leads to 10-dimensional superstrings of type IIA,
with the M5 branes turned into D4 branes. This leads to a nonconformal duality
of a 5-dimensional supersymmetric field theory to a string theory on a background that is related
to AdS$_6\times S^4$ by a Weyl transformation \cite{Kanitscheider:2008kd}. 
Witten \cite{Witten:1998zw}
suggested to consider one further dimensional reduction on a circle for the superfluous
fourth spatial coordinate, 
\begin{equation}
 x_4 + 2\pi R_4\equiv x_4 + 2\pi M_{\rm KK}^{-1} \simeq x_4,
\end{equation}
but with
supersymmetry breaking boundary conditions for fermions as they also appear in
finite temperature field theory in the imaginary-time formalism.
All modes not protected by gauge symmetry become massive in this case, fermionic gluinos at tree-level, and 
adjoint scalars through quantum effects, leaving pure-glue Yang-Mills
theory as low-energy effective theory. The 5-dimensional supersymmetric field theory thus turns into
ordinary 4-dimensional Yang-Mills theory. On the other hand, the background geometry
of the string theory becomes
\be\label{ds2W}
 ds^2=\left(\frac{u}{R}\right)^{3/2}\left[\eta_{\mu\nu}dx^\mu dx^\nu+f(u)dx_4^2\right]+\left(\frac{R}{u}\right)^{3/2}\left[\frac{du^2}{f(u)}+u^2 d\Omega_4^2\right]
\ee
with
\be
f(u)=1-\frac{\uKK^3}{u^3},\quad \MKK=\frac32 \frac{\uKK^{1/2}}{R^{3/2}}
\ee
and a nonconstant (``linear''\footnote{Since it would be linear in suitable radial coordinates.}) dilaton $e^\phi=g_s(u/R)^{3/4}$.
Here $u\ge \uKK$ is the holographic direction with $u=\infty$ corresponding to the conformal boundary. The topology of the subspace formed by $u$ and the Kaluza-Klein direction $x_4$
is that of a cigar with the $x_4$ circle shrinking to zero size at $u=\uKK$. This 
lower bound on $u$ corresponds to the would-be horizon of
a Euclidean black-hole geometry and the relation between $\uKK$ and $\MKK$
is determined by the absence of a conical singularity at $u=\uKK$.

The parameters of the 4-dimensional gauge theory with Lagrangian $\mathcal L=-\frac1{2g^2}\Tr F_{\mu\nu}F^{\mu\nu}$ resulting from the reduction of $\mathcal L=-\frac1{2g_5^2}\Tr F_{MN}F^{MN}+\ldots$ with $g_5^2=2g_s (2\pi)^2 \ell_s$
are given by\footnote{\label{fng}In virtually all of the literature on the Witten(-Sakai-Sugimoto) model, $\mathcal L=-(4g_{\rm YM}^2)^{-1}\Tr F_{\mu\nu}F^{\mu\nu}$
is used. This different convention has no consequences as long as no comparison
to perturbative QCD is made, where the standard definition of the SU(3) gauge coupling is $g$ with
$g^2=2g_{YM}^2$ and the $\alpha_s$ of QCD reads $\alpha_s=g^2/4\pi=g_{YM}^2/2\pi$.}
\be
g^2=\frac{g_5^2}{2\pi R_4}=4\pi g_s\ell_s\MKK,\quad \frac{R^3}{\ell_s^3}=\pi g_s N_c,
\ee
where the latter equation is due to the nonvanishing 
Ramond tensor field sourced by the $N_c$ D4-branes. 
For the supergravity approximation to be justified, the maximum curvature %, which occurs at $u=\uKK$
$\sim(\uKK R^3)^{-1/2}$ located at $u=\uKK$
needs to be small compared to the string tension $\propto \ell_s^{-2}$,
which again requires large 't Hooft coupling $\lambda=g^2N_c\gg 1$.

The dimensionless coupling $g$ is derived from the dimensionful coupling $g_5$ at the Kaluza-Klein
mass scale $\MKK$, beyond which the field theory reveals its underlying 5-dimensional nature.
Ideally, $\MKK$ should be sent to infinity, 
but that would mean small 't Hooft coupling, where string theory
effects beyond the supergravity approximation would be needed. In the Witten model, we have to be content with the strong-coupling
regime below some finite scale $\MKK$, but the fact that this dual theory may be 
continuously connected to real QCD makes it an
interesting approximation to the strong-coupling regime of the latter.

A particularly simple quantity to calculate in the Witten model is the Wilson
loop of heavy (nondynamical) quarks. Connecting two heavy quarks at the boundary which are
widely separated along $x$ through a Wilson line is represented by
a fundamental string embedded in the bulk geometry. 
The string tension is locally given by $(2\pi\alpha')^{-1}\sqrt{-g_{tt}g_{xx}}$.
For large separation in $x$, the
energetically favored solution is clearly 
to have as much as possible of the length of the string
at the minimal value of
$\sqrt{-g_{tt}g_{xx}}$, which occurs at $u=\uKK$. The effective string tension at
large $\Delta x$ thus approaches a constant which indicates confinement of quarks
and which is
given explicitly by
\be\label{sigmastring}
\sigma=\frac1{2\pi\ell_s^2}\sqrt{-g_{tt}g_{xx}}\Big|_{u=\uKK}=
\frac1{2\pi\ell_s^2}\left(\frac{\uKK}{R}\right)^{3/2}=\frac{g^2 N_c}{27\pi}\MKK^2.
\ee

In accordance with confinement, there is a mass gap in the spectrum
of fluctuations of the background geometry %in the Witten model 
which is proportional to $\MKK$.
These fluctuations can be classified with respect to their 4-dimensional
quantum numbers and comprise scalar, vector, and tensor modes, which correspond to
glueballs in the gauge theory.

\begin{figure}[t]
\centerline{\includegraphics[width=0.7\textwidth,height=0.5\textwidth]{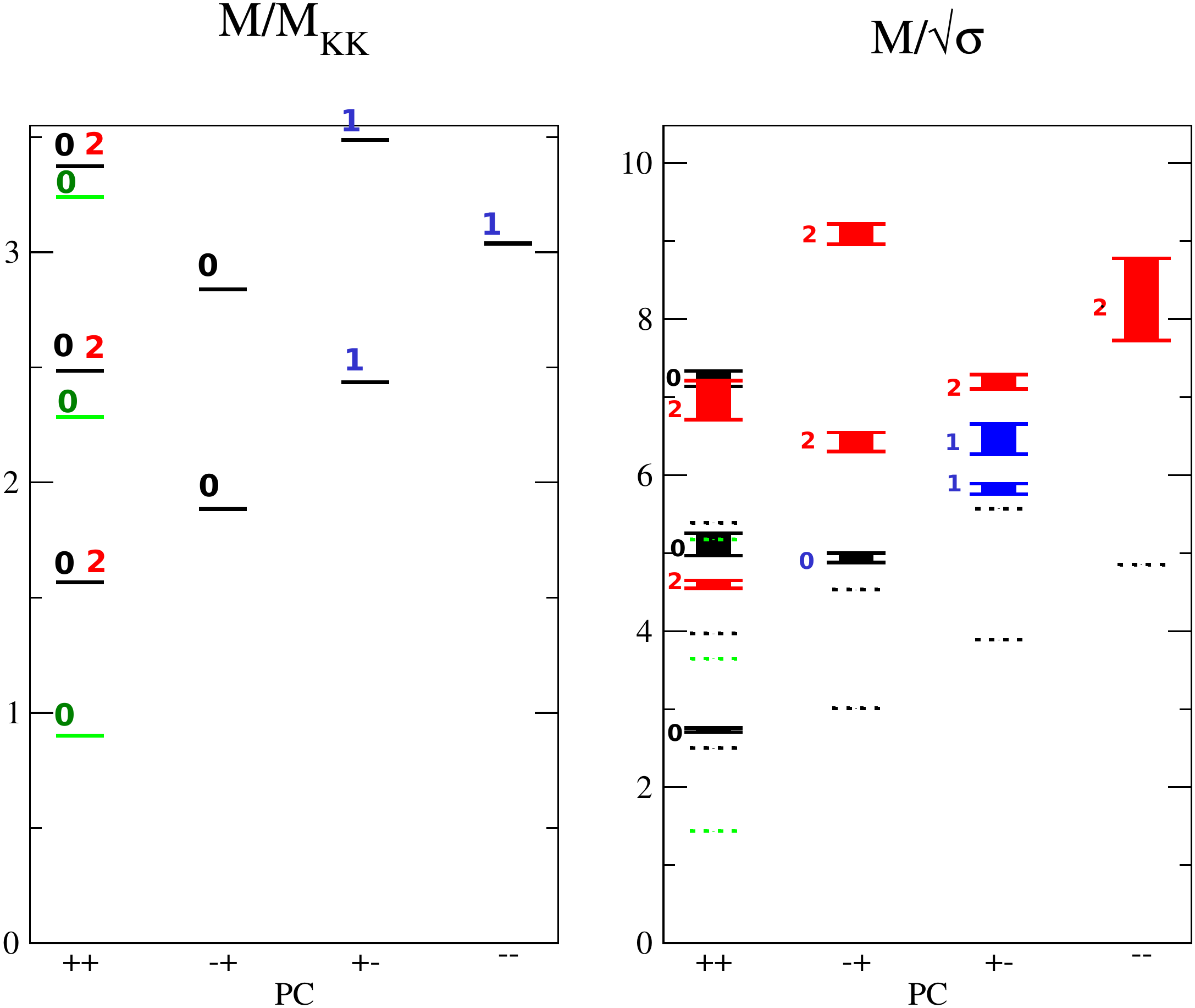}} %{gbspectrumsigma2}} %{figspectrum.png}
\centerline{\small \hfil (a) \hfil\hfil (b) \hfil}
\medskip
\caption{Glueball spectrum of the Witten model (a) as obtained
in Ref.~\cite{Brower:2000rp}, in units of $\MKK$, (``exotic'' modes in green), compared to
the recent large-$N_c$ lattice results of Ref.~\cite{Lucini:2010nv} (b) in
units of $\sqrt{\sigma}$ but with overall scale such that the lowest tensor mode
are on equal height. The dotted lines in (b) give the holographic spectrum
in terms of the holographic string tension with the standard parameter set of
the Sakai-Sugimoto model (\ref{kappaMSS}).}
\label{figGB}       % Give a unique label
\end{figure}

Following up on pioneering work in Ref.~\cite{Csaki:1998qr,%de Mello Koch:1998qs,
Csaki:1999vb,Constable:1999gb}, the complete spectrum was
worked out in Ref.~\cite{Brower:2000rp}. In Fig.~\ref{figGB}, the spectrum
in the Witten model (Fig.~\ref{figGB}a) is compared with recent lattice
results \cite{Lucini:2010nv} for large-$N_c$ Yang-Mills theory (colored boxes in Fig.~\ref{figGB}b)
with the vertical scale chosen such that the lowest tensor glueball $2^{++}$
in the two figures have equal height. The holographic calculation appears to
reproduce qualitatively many of the conspicuous features of the lattice result
which is in fact rather similar to those obtained at $N_c=3$. The main qualitative difference
is the absence of glueballs with quantum numbers $2^{-+}$ and of glueballs with spin $>2$
in the Witten model. Another peculiar feature of the holographic result is
a certain proliferation of scalar modes $0^{++}$: the green lines in Fig.~\ref{figGB}a
represent scalar modes which have a polarization involving the metric component
$g_{44}$ and which have been termed ``exotic'' in Ref.~\cite{Constable:1999gb}. The other
modes in the $PC=++$ sector are degenerate $0^{++}$ and $2^{++}$ glueballs.
Thus there are 4 scalar glueballs with mass less or equal the first excited
tensor glueballs, whereas the lattice features only two in that range.
In fact, Ref.~\cite{Constable:1999gb} already suspected that the scalar glueballs
associated with the ``exotic'' polarization may perhaps not correspond to
glueballs in QCD, but later Ref.~\cite{Brower:2000rp} found that the lowest
exotic mode is the lowest-lying one of all the modes, making the overall
picture tantalizingly similar to that found in lattice simulations.
Below we shall revisit this issue when quarks are included by the
extension provided by the Witten-Sakai-Sugimoto model.

In Ref.~\cite{Kanitscheider:2008kd} the necessary tools for holographic
renormalization for nonconformal holography have been developed and applied
to the Witten model to calculate the gluon condensate with the result
\be
\frac14 \left<\Tr F^2\right>=\frac{2N_c}{3^7\pi^2}(g^2 N_c/2)^2\MKK^4,
\ee
where we have taken into account that the $g_4$ used in Ref.~\cite{Kanitscheider:2008kd}
differs from the standard definition of $g$ of perturbative QCD by $g_4^2=g^2/2$.
%In principle, this can be used to fix the parameters  %of the Witten model
The gluon condensate as usually defined in QCD 
in terms of
canonically normalized gluon fields $G_{\mu\nu}^a$
is given by
% is usually defined in terms of
% canonically normalized gluon fields so that
\be\label{gluoncondensate}
C^4\equiv\left<\frac{\alpha_s}{\pi}G_{\mu\nu}^a
G^{a\mu\nu}\right>=\frac1{2\pi^2}\left<\Tr F^2\right>=
\frac{4N_c}{3^7\pi^4}(g^2 N_c/2)^2\MKK^4.
\ee
We shall come back to this when discussing the choice of parameters of the Sakai-Sugimoto model.

\subsection{Inclusion of chiral quarks: the Sakai-Sugimoto model}

Type-IIA string theory has stable D-branes with any even number of spatial dimensions.
If such additional branes are present, the open strings connecting those with the color D4-branes
carry one color index on one end and a new index on the other, which
can represent the flavor quantum number of quarks. If flavor and color branes intersect,
one gets chiral quarks. As long as the number $N_f$ of these additional branes is small compared to $N_c$,
they can be considered as ``probes'' with negligible back-reaction on the background
geometry they are imbedded in. This is analogous to the introduction of quenched
quarks in lattice QCD.

A first interesting attempt to include quarks in the Witten model was
made by Kruczenski, Mateos, Myers, and Winters \cite{Kruczenski:2003uq} by
introducing D6 flavor branes in the background of the Witten model.
In the case of only one flavor brane and a massless quark, 
a spontaneous
breaking of an axial U(1) symmetry was observed, but the generalization to
a nonabelian flavor symmetry at $N_f>1$ did not give a good model for chiral
symmetry breaking. 

\begin{figure}[h]
\centerline{\includegraphics[width=0.4\textwidth]{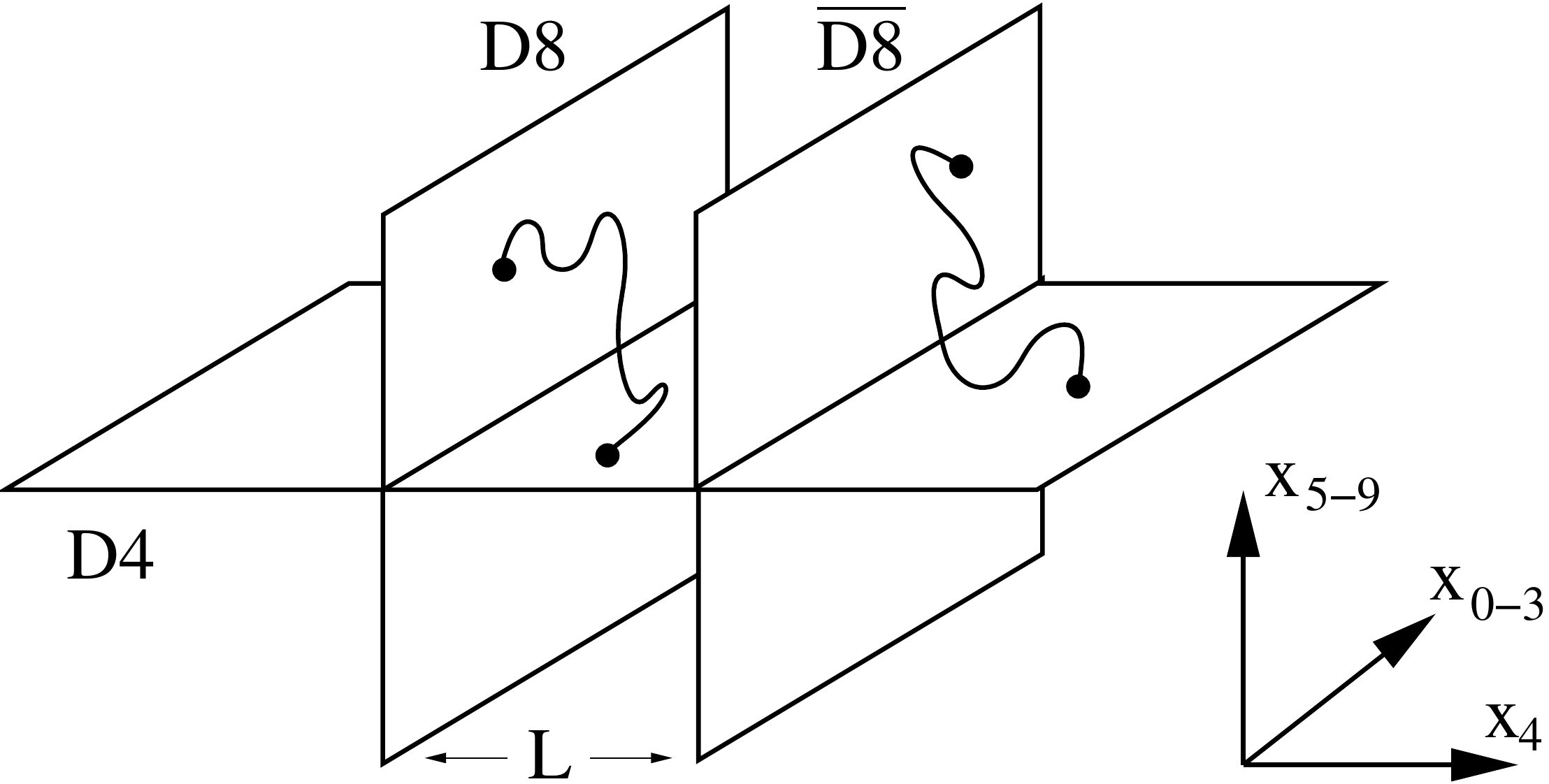}\qquad
\includegraphics[width=0.35\textwidth]{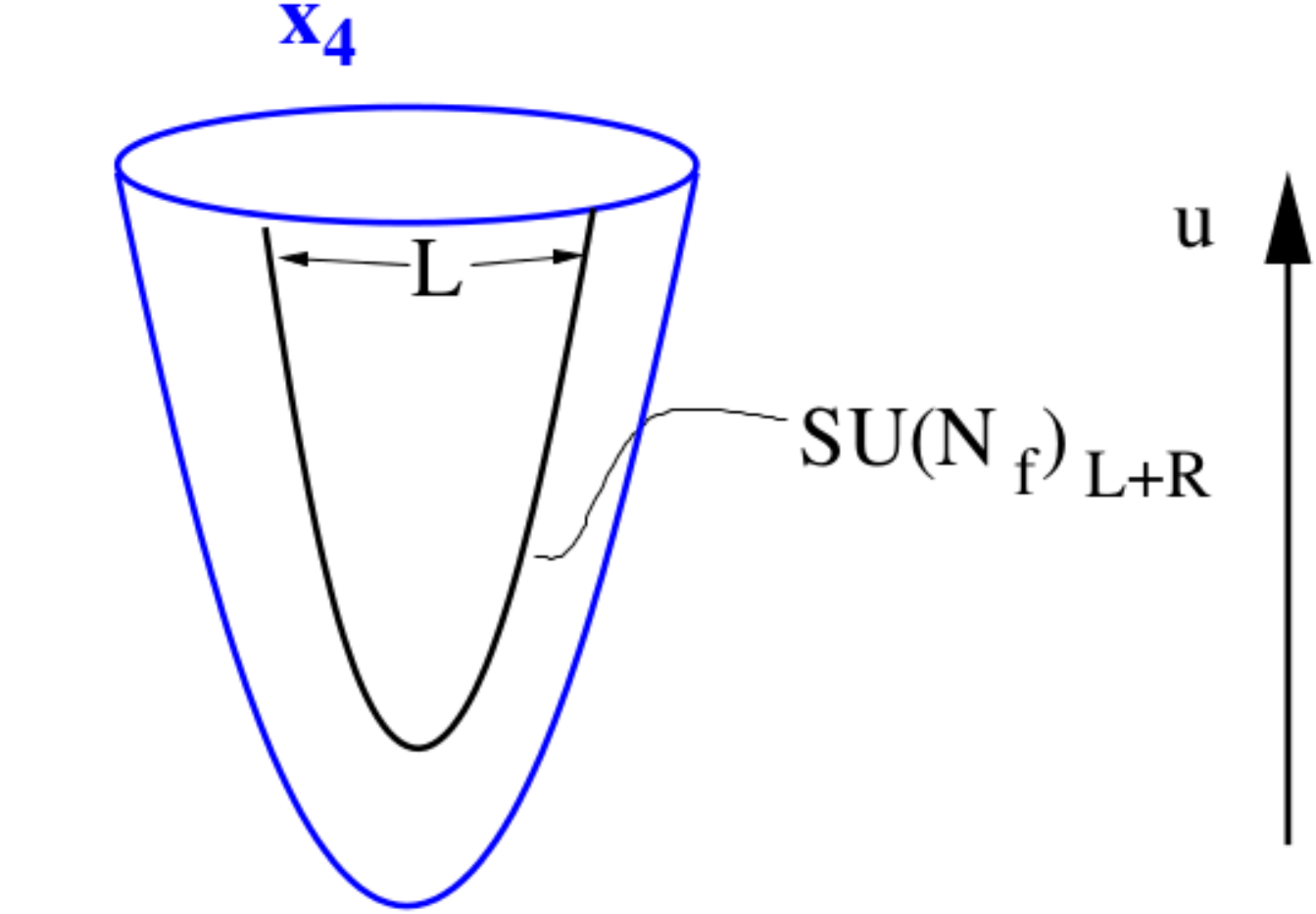}}
\centerline{\small \hfil (a) \hfil\hfil (b) \hfil}
\medskip
\caption{(a) D4-D8-$\overline{\mbox{D8}}$ brane configuration in the 10 dimensions $x_0,\ldots,x_9$,
with $x_4$ understood as periodic; (b) near-horizon geometry with $u$ being a radial coordinate in
the transverse space $x_{5\ldots 9}$.}
\label{figD8}       % Give a unique label
\end{figure}

This problem was solved by Sakai and Sugimoto \cite{Sakai:2004cn,Sakai:2005yt} by instead using
pairs of probe\footnote{The daunting issue of backreaction of localized D8 branes has been
studied in Ref.~\cite{Burrington:2007qd}. A different strategy to include dynamical
quarks has been recently proposed in Ref.~\cite{Bigazzi:2014qsa} through smeared
D8 branes and corrections to first order in $N_f/N_c$ have been calculated.
A price of the latter approach is however that the global flavor symmetry is
reduced to $\mathrm U(1)^{N_f}_L\times \mathrm U(1)^{N_f}_R$.}
D8 and anti-D8 branes filling all spatial directions except the $x_4$ (Fig.~\ref{figD8}a).
This introduces chiral quarks and anti-quarks localized at different points
of the Kaluza-Klein circle. The global flavor symmetry $\mathrm U(N_f)_L\times \mathrm U(N_f)_R$
of the field theory living on the boundary
corresponds to a local gauge symmetry on the flavor branes, which is broken
spontaneously to the diagonal subgroup because the D8 and anti-D8 branes have
nowhere to end except by joining in the bulk (Fig.~\ref{figD8}b).

The spectrum of fluctuations of the flavor gauge fields on the D8 branes include
a zero mode for the fifth component $A_u(x^\mu,u)$ whose gauge-invariant holonomy
represents the Goldstone bosons of the spontaneously broken $\mathrm U_A(N_f)$:
\be
U(x^\mu)={\rm P}\,\exp\left\{ i \int_{-\infty}^\infty dZ' A_Z(x^\mu,Z)\right\}\equiv
\exp\left\{ 2i\pi(x^\mu)/f_\pi \right\} \in \mathrm U(N_f),
\ee
where we have switched to a new dimensionless 
coordinate $Z$ defined by $(u/\uKK)^3=1+Z^2$ %1+(z/\uKK)^2$
which runs from $-\infty$ to $+\infty$ as one follows the holographic direction
from the boundary into the bulk along the D8 brane and back along the anti-D8 brane.
Here we have assumed that the D8 and the anti-D8 brane are maximally separated
so that they connect at the tip of the cigar, $u=\uKK$.

In massless QCD, the axial U(1)$_A$ symmetry is broken by an anomaly, which
is suppressed at large $N_c$. However, this effect has also a beautiful supergravity description 
\cite{Barbon:2004dq} involving the Ramond-Ramond 1-form whose flux through the
surface parametrized by $x_4$ and $u$ is related to the $\theta$ parameter.
The mass of the $\eta'$ pseudoscalar meson is related by the Witten-Veneziano
formula to the topological susceptibility, and was calculated by
Sakai and Sugimoto  \cite{Sakai:2004cn} as\footnote{Note that Ref.~\cite{Sakai:2004cn}
uses $g_{\rm YM}$ and that our $g^2=2g_{\rm YM}^2$.}
\be\label{metaprime}
m_{\eta'}=\frac1{3\sqrt3 \pi}\sqrt{\frac{N_f}{N_c}}(g^2 N_c/2)\MKK.
\ee

To leading order in $N_c$ and $\alpha'$, the D8-brane action is given explicitly by
\be
S_{\rm D8}=-\kappa\int d^4x\,dZ\, \Tr\left[
\frac12 (1+Z^2)^{-1/3}F_{\mu\nu}^2+(1+Z^2)\MKK^2 F_{\mu Z}^2+O(\alpha'{}^2 F^4)
\right]+S_{\rm CS},
\quad\kappa=\frac{(g^2N_c/2)N_c}{216\pi^3},
\ee
where all fields have been taken as constant with respect to the $S^4$ whose
volume has been integrated over. 
$S_{\rm CS}$ is the unavoidable Chern-Simons action of D-branes which here
can be shown to reproduce the correct chiral anomaly of QCD, and also
the correct Wess-Zumino-Witten term in the chiral Lagrangian \cite{Sakai:2004cn}. The well-known
5-dimensional formulation of the latter thus appears in a most natural way;
background gauge fields for the chiral symmetry can be introduced as
nontrivial boundary conditions $A^{L,R}_\mu(x)=\lim_{Z\to\pm\infty} A_\mu(Z,x)$.

The Goldstone bosons and the $\eta'$ meson
are contained in $A_Z=\phi_0(Z)\pi(x)$ with $\phi_0\propto (1+Z^2)^{-1}$.
Requiring a canonical normalization of the kinetic term for the Goldstone bosons in
\be\label{SD80}
S_{\rm D8}=\frac{f_\pi^2}{4}\int d^4x\, \Tr\left( U^{-1}\6_\mu U \right)^2+\ldots,
\ee
fixes the so-called pion decay constant in terms of $\lambda$ and $\MKK$ as
\be
f_\pi^2=\frac1{54\pi^4}(g^2N_c/2)N_c\MKK^2.
\ee

Vector and axial vector mesons appear as even and odd 
eigenmodes of $A_\mu^{(n)}=\psi_n(Z) v^{(n)}_\mu(x)$
with equation
\be\label{psin}
-(1+Z^2)^{1/3}\6_Z\left( (1+Z^2)\6_Z \psi_n \right)=\lambda_n \psi_n,\quad
\psi_n(\pm\infty)=0.
\ee
The lowest mode is interpreted as the vector meson $\rho$ with
mass $m_\rho^2=\lambda_1 \MKK^2$ and $\lambda_1=0.669$.
The next mode with $\lambda_2=1.569$ is an axial vector meson that
can be identified \cite{Sakai:2004cn} with the $a_1(1260)$.
The experimental value $m^2_{a_1(1260)}/m^2_\rho\approx 2.52$ is
surprisingly close to $\lambda_2/\lambda_1\approx 2.344$.
Also the ratio of the squared mass of the next vector meson, $\rho(1450)$,
to that of the $\rho$ meson, $m^2_{\rho(1450)}/m^2_\rho\approx 3.57$ compares
quite well with $\lambda_3/\lambda_1\approx 4.29$. However, this
nice agreement may be fortuitous. Recent lattice simulations of the spectrum
of mesons in large-$N_c$ QCD extrapolated to
zero quark masses have obtained larger values \cite{Bali:2013kia}:
$m^2_{a_1}/m^2_{\rho}\approx 3.46$ and $m^2_{\rho^*}/m^2_{\rho}\approx 5.77$.
This would corresond to an error of 21\% and 16\%, respectively, for the 
mass ratios (unsquared). Given that the Sakai-Sugimoto model is bound to
show deviations from QCD above the scale of $\MKK\approx 1.22%232 
\,m_\rho$, this
may still be taken as a notable success.

In Ref.~\cite{Sakai:2004cn,Sakai:2005yt} the physical values of 
$m_\rho\approx 776$~MeV and $f_\pi\approx 92.4$~MeV have been used to set the free parameters
\be\label{kappaMSS}
\MKK=949\,{\rm MeV}, \qquad  \kappa=7.45\cdot 10^{-3} \;\Rightarrow\; \lSS\equiv g^2N_c/2\approx 16.63,
\ee
%the latter corresponding to $g^2N_c/2\approx 16.63\equiv\lSS$ 
($\lSS=\gYM^2 N_c$ is the 't Hooft coupling %$\lambda$
as used in Ref.~\cite{Sakai:2004cn,Sakai:2005yt}; cf.~footnote \ref{fng}), and these values have been used widely since.
At $N_c=3$, this amounts to $\alpha_s(\MKK)\approx 0.88$, which is of the
order of magnitude to expect from perturbative QCD at such low scales, albeit somewhat
high when compared to the particular values obtained in the $\overline{\rm MS}$-scheme,
$\bar\alpha_s(1\,{\rm GeV})\approx 0.5$. But one should keep in mind that
$\bar\alpha_s$ grows sharply at 1~GeV with large renormalization scheme dependence.
A better quantity to consider is the renormalization-scheme independent string tension.
The holographic result (\ref{sigmastring}) depends on
both $\MKK$ and $\lSS$, which yields with the above choice (\ref{kappaMSS})
$\sqrt{\sigma}=0.6262\MKK\approx 594$~MeV.
This is higher than the value  $\sqrt{\sigma}\sim 440$~MeV extracted from
lattice calculations \cite{Teper:1997am}, also suggesting
that $\bar\alpha_s$ is somewhat too large, 
since in (\ref{sigmastring}) $\sigma\propto\alpha_s$.

% Since the Sakai-Sugimoto model involves chiral quarks, it may be
% more appropriate to use $f_\pi$ in the chiral limit as input parameter.
% Lattice results at $N_c=3$ \cite{Colangelo:2010et} indicate that
% $f_\pi/f_\pi^{\rm chiral}=1.073(15)$, whereas at large $N_c$ this
% goes down to \cite{Bali:2013kia} $1.020(20)$.

% With $\MKK$ and $\lSS$ fixed, the Sakai-Sugimoto model pins down all the results
% of the Witten model. 
Another quantity of immediate interest which depends on both $\MKK$ and $\lSS$
is the gluon condensate.
(The glueball spectrum, which only depends on $\MKK$, 
will be revisited in the next section.)
%The gluon condensate (\ref{gluoncondensate}) depends on both $\MKK$ and $\lSS$.
% With $\MKK$ fixed by the mass of the $\rho$ meson,
% one could in principle also use the result (\ref{gluoncondensate}) for the gluon condensate
% to fix the coupling $\lSS$ at $\MKK$. 
Inserting the parameters (\ref{kappaMSS}) and $N_c=3$
yields 
\be\label{C4SS}
C^4=0.0126\,{\rm GeV}^4\quad (\lSS=16.63),
\ee
which happens to almost coincide with
the standard SVZ sum rule \cite{Shifman:1978bx} value $C^4=0.012\,{\rm GeV}^4$ in QCD. 
% However
% Ref.~\cite{Zyablyuk:2002kg} finds $C^4<0.008\,{\rm GeV}^4$,
% Ref.~\cite{Samsonov:2004zm,Ioffe:2005ym} find $C^4\approx 0.005\,{\rm GeV}^4$.
However, both significantly smaller \cite{Ioffe:2005ym,Samsonov:2004zm} and larger
values \cite{Narison:2011xe} have been obtained using sum rules. Lattice simulations
typically give significantly larger values, but of the same size as ambiguities from
the subtraction procedure \cite{Bali:2014sja}. Thus it appears that the gluon
condensate is not very suited to really
determine the 't Hooft coupling in the Witten(-Sakai-Sugimoto)
model. A moderate value of the gluon condensate may however be considered
to be a prediction of this model.

Using the result for the string tension, one can also compare the prediction
for the dimensionless ratio
\be
m_\rho/\sqrt{\sigma}\approx 1.306\quad (\lSS=16.63)
\ee
with the large-$N_c$ lattice
result of Ref.~\cite{Bali:2013kia}, which reads 1.504(50).
This agrees within 15\%, and the fact that the holographic result is smaller
also suggests that the assumed value for $\lSS$ and thus $\sigma$ in the Sakai-Sugimoto model
is perhaps too large. Turning this around, and fitting the lattice result for $m_\rho/\sqrt{\sigma}$
would give $\lSS\approx 12.55(80)$ resulting in more moderate values
$\sqrt{\sigma}\approx 515$~MeV and $\alpha_s(\MKK)\approx 0.66$.
% At $N_c=3$, the lattice value for $m_\rho/\sqrt{\sigma}$ increases to $\approx 1.75$. [Ref?] Using this
% as an input would correspond to $\sqrt{\sigma}\approx 443$ ~MeV and $\alpha_s(\MKK)\approx 0.49$.
% % expressed in terms of the string tension, as done in the large-$N_c$ lattice
% study of Ref.~\cite{Bali:2013kia}. Doing so, one finds that
% the value $m_\rho/\sqrt{\sigma}$ predicted by the Sakai-Sugimoto model
% agrees with the lattice result within 18\%. 
%, while the deviations for
% the $a_1$ and the $\rho^*$ are larger: 43\% and 37\%, respectively.
This also corresponds to a 13\% lower value of the pion decay constant, $f_\pi=80.3$~MeV, and one
might indeed argue in favor of a reduced $f_\pi$ because the Sakai-Sugimoto model is strictly chiral
(although, judging from the lattice results of Ref.~\cite{Colangelo:2010et}, $f_\pi$ should have a
reduction around 7\% for $N_c=3$, 
whereas at large $N_c$ Ref.~\cite{Bali:2013kia} found only a 2\% reduction).
At any rate, a downward variation of $\lSS$ from 16.63, say to 12.55, should give a reasonable
theoretical error bar for quantitative predictions of the Witten-Sakai-Sugimoto model.

In Ref.~\cite{Imoto:2010ef}, Imoto, Sakai, and Sugimoto have more recently studied the predictions
of the Sakai-Sugimoto model for other mesons, which are not covered by the flavor gauge fields
on the D8-branes. The latter correspond to massless open string modes
and only give the Goldstone bosons and a tower of $1^{--}$ and $1^{++}$ mesons.%
\footnote{Originally, also excitations of the brane embedding function were considered in Ref.~\cite{Sakai:2004cn}
and interpreted as a tower of $0^{++}$ and $0^{--}$ scalar mesons. These are, however, odd under
the so-called \cite{Brower:2000rp} $\tau$-parity, and should be discarded since
the quarks and gluon are invariant under this discrete symmetry \cite{Imoto:2010ef}.
It is therefore not disturbing that the prediction for the mass of the $a_0(1450)$ meson in the original paper
of Sakai and Sugimoto does not agree well with the recent large-$N_c$ lattice
result of Ref.~\cite{Bali:2013kia}.}
Despite the fact that gauge/gravity duality works in the decoupling limit corresponding to
$\ell_s\to0$, the masses of the massive string modes remain finite, and yield a host of
mesonic states. 
%, showing that the duality of the Witten-Sakai-Sugimoto model is one that involves
% string aspects at its core. 
The spectrum of those further states
is very difficult to evaluate precisely because of the
curved background. Ref.~\cite{Imoto:2010ef} succeeded, however, in calculating corrections
to the Regge trajectory of the $\rho$-meson and could reproduce the required curvature
of the trajectory around zero mass to match with a Regge intercept $\alpha_0=1$.
In order to match the experimental values of the meson masses
on the $\rho$-trajectory, Ref.~\cite{Imoto:2010ef} again observed that the model needs
a lower value of the string tension than given by the standard choice of parameters.
Their fit even points to a rather low value of $\sqrt{\sigma}=380$~MeV. Curiously enough, also
the large-$N_c$ lattice study of Ref.~\cite{Bali:2013kia} found a reduced value of $\sqrt{\sigma}$, in their case $395$~MeV,
to give a better fit than $440$~MeV.

The Sakai-Sugimoto model is not limited to the meson sector of low-energy QCD.
The D8-brane action (\ref{SD80}), restricted to the Goldstone bosons and integrated over the
holographic coordinate, reads, to leading order in 't Hooft coupling and $N_c$:
\be\label{SD8skyrme}
S_{\rm D8}=\int d^4x\left\{ \frac{f_\pi^2}{4}\Tr\left( U^{-1}\6_\mu U \right)^2+
\frac{1}{32e^2}\Tr[U^{-1}\6_\mu U,U^{-1}\6_\nu U]^2 \right\},\quad e^{-2}=\frac{15.253}{27\pi^7}
g^2 N_c^2,
\ee
with $f_\pi^2\propto g^2 N_c^2$ given already in (\ref{SD80}). This coincides with the Skyrme
model \cite{Schechter:1999hg} with a definite prediction of the Skyrme parameter $e$
and so allows baryons to appear as solitons of
the chiral Lagrangian. In the bulk, the corresponding object is a holographic baryon vertex
\cite{Witten:1998xy} in the form of additional D4-branes wrapping the $S^4$ in the 10-dimensional background geometry.
Because of the RR flux on the $S^4$ the D4-brane is connected to the D8 branes by $N_c$ open strings,
which pull it inside the D8 branes where it appears as an instanton configuration, realizing
a connection between instantons in 5-dimensional gauge theory and the Skyrmion recognized
prior to the Sakai-Sugimoto model in Ref.~\cite{Son:2003et}.

A full-fledged treatment of baryons in the Sakai-Sugimoto model
has proved to be very interesting but difficult \cite{Hong:2007kx,Hata:2007mb,Hashimoto:2008zw,Seki:2008mu},
suggesting that string corrections should be taken into account. 
Considering the baryon mass, it should be given to leading order by the mass of a
D4-brane at the tip of the D8 brane configuration, which is \cite{Sakai:2004cn}
%\be
$m_{\rm D4}=8\pi^2\kappa\MKK=\frac1{27\pi}(g^2 N_c/2)N_c\MKK.$
%\ee
In \cite{Hong:2007kx,Hata:2007mb} the correction coming from the U(1)$_V$-field induced on
the D8-brane and leading to a contribution from $S_{\rm CS}$ has been calculated with the result
\cite{Hong:2007kx,Hata:2007mb}
\be
M_B=\left[\frac1{27\pi}(g^2 N_c/2)+\sqrt{\frac{2}{15}} + O((g^2N)^{-1})\right] N_c\MKK.
\ee
With the standard parameter choice this corresponds to $(558+1040=1598)$~MeV, which is too high
when compared to real nucleons. But the lack of convergence (the correction term is about
twice the leading term) seems to indicate that an expansion about a pointlike instanton
is problematic, at least for the 't Hooft coupling required to make contact with QCD.
Curiously enough,
if one just considers the effective Skyrme action (\ref{SD8skyrme})
and evaluates the well-know result for the energy of a Skyrmion \cite{Adkins:1983ya,Schechter:1999hg}, 
$E\approx 23.2\pi f_\pi/e$,
with the above parameters, one obtains $M_N\approx 920$~MeV, 
tantalizingly close to the real-world value. This may be taken as an indication that
one need not lower the scale $\MKK$ to make contact with QCD (as suggested in \cite{Hata:2007mb}), but should rather
aim at a different approximation scheme for holographic baryons.

As another extreme we can also consider the mass of the $\eta'$-meson given in (\ref{metaprime}),
which is proportional to $\sqrt{N_f/N_c}$
and thus parametrically suppressed at large $N_c$. However, extrapolating to
$N_c=3$ and inserting again
the above parameters for $\lSS$ and $\MKK$ this yields
$m_{\eta'}\approx 967$~MeV for $N_f=3$, astonishingly close to the experimental value of 958~MeV.
With the alternative choice $\lSS=12.55$, the lower value of 730~MeV is obtained, which is
seemingly disfavored, but may actually be more appropriate because of the chiral limit
and the missing strange quark mass.

In Table \ref{tabparams}, a summary of the numerical predictions
are given, including for completeness a variation of the Sakai-Sugimoto model
where the D8 branes are not maximally separated \cite{Antonyan:2006vw,Aharony:2006da}. 
They then join at
a point $u_0>\uKK$, and one may interpret the difference between $u_0$ and $\uKK$ as
a constituent quark mass, given by the mass of a string stretched from $u_0$ ot $\uKK$. In Ref.~\cite{Callebaut:2011ab} the latter has been
set to 310~MeV, which after matching the $\rho$ meson mass leads to a significantly smaller
value of $\MKK$ and a slighter smaller value of $\lSS$. As remarked in Ref.~\cite{Callebaut:2011ab},
this leads to a better match of the string tension as observed on the lattice. On the other
hand, the gluon condensate becomes uncomfortably small. Also the other meson
masses fit somewhat less well.

To wrap up this numerological discussion, % (summarized in Table \ref{tabparams}), 
the original choice of parameters of the Sakai-Sugimoto model
seems to be in quite good shape, but one may want to consider also 
smaller 't Hooft couplings in view of the somewhat high string tension.
At any rate, the fact that most of the quantities considered here
come out in the right ballpark is certainly remarkable given that this model has a
cutoff scale of $\MKK\sim 1$~GeV. A reason may be that this cutoff is very smooth in comparison 
with a sharp momentum or lattice cutoff
and that it also fully respects 3+1-dimensional Lorentz symmetry.

\begin{table}
\centering
\caption{Choice of parameters of the Witten-Sakai-Sugimoto model and resulting predictions in 3 variants: (i) Original choice of Sakai and Sugimoto
\cite{Sakai:2005yt} which fits $\MKK$ to the experimental value of $m_\rho$ and
the 't Hooft coupling $\lSS$ through the experimental value of $f_\pi$; (ii) same, except that
$\lSS$ is fitted such as to reproduce the large-$N_c$ lattice result of
Ref.~\cite{Bali:2013kia} for $m_\rho/\sqrt{\sigma}\approx 1.504$; (iii) non-antipodal D8 brane configuration with $u_0/\uKK\approx 1.38$ such as to have a constituent quark mass of $m_q^{\rm constit.}=310$~MeV \cite{Callebaut:2011ab}. (Starred numbers are directly fixed by the input parameters.)
\label{tabparams}}       % Give a unique label
% For LaTeX tables you can use
\begin{tabular}{l|lll}
%\hline
~& (i) antipodal \cite{Sakai:2005yt}&  (ii) antipodal & (iii) non-antipodal \cite{Callebaut:2011ab}\\
fitted to: & ($m_\rho,f_\pi$) & ($m_\rho,m_\rho/\sqrt{\sigma}$) & ($m_\rho,f_\pi,m_q^{\rm constit.}$) \\
$\MKK$ [MeV] & 949 & 949 & 720.1 \\
$\lSS$ ($=6\pi\alpha_s$) & 16.63 & 12.55 & 15.13 \\
\hline
$f_\pi$ [MeV] & 92.4$^*$ & 80.3 & 93$^*$ \\
$m_\rho$ [MeV] & 776$^*$ & 776$^*$ & 776$^*$ \\
$\sqrt{\sigma}$ [MeV] & 594 & 515$^*$ & 430 \\
$C^4$ [GeV$^4$] & 0.0126 & 0.0072 & 0.0035 \\
$m_{\eta'}|_{N_f=3}$ [MeV] & 967 & 730 & 667 \\
%$\lambda_2/\lambda_1$ & 2.344 & 2.344 & 2.182 \\
$m_{a_1(1260)}$ [MeV] & 1188 & 1188 & 1146 \\
$M_G$ [MeV] & 855 & 855 & 649 \\
$M_{D,T}$ [MeV] & 1487 & 1487 & 1128 \\
\hline
\end{tabular}
\bigskip
% Or use
%\vspace*{5cm}  % with the correct table height
\end{table}

\begin{table}
\centering
\caption{Holographic glueball masses ($\MKK=949$~MeV). The glueball candidate
corresponding to the ``exotic'' polarization of the metric field is marked by ``(E)''.}
\label{tab-1}       % Give a unique label
% For LaTeX tables you can use
\begin{tabular}{l|lr}
%\hline
$J^{PC}$ & $M/\MKK$ %& $M/\sqrt{\sigma}$ 
& $M$[MeV] \\\hline
$0^{++}$ (E)& 0.90113 & 855 \\
$0^{++}$-$2^{++}$ & 1.56690 & 1487 \\
$0^{-+}$ & 1.88518 & 1789 \\
$0^{*++}$ (E)& 2.28487 & 2168 \\
$1^{+-}$ & 2.43529 & 2311 \\
$0^{*++}$-$2^{*++}$ & 2.48514 & 2358 \\
$0^{*-+}$ & 2.83783 & 2693 \\
$1^{--}$ & 3.03763 & 2883 \\
\hline
\end{tabular}
\bigskip
% Or use
%\vspace*{5cm}  % with the correct table height
\end{table}

\begin{table}
\centering
\caption{The masses of isotriplet mesons given by the modes
of the flavor gauge field on the (maximally separated) D8 branes, compared to the experimental value $(m/m_\rho)^{\rm exp.}$ from [PDG] ($\rho(770),a_1(1260),\rho(1450),a_1(1640)$) and to the large-$N_c$ results from Ref.~\cite{Bali:2013kia}.}
\label{tab-2}       % Give a unique label
% For LaTeX tables you can use
\begin{tabular}{l|llll}
%\hline
Isotriplet Meson & $\lambda_n=m^2/\MKK^2$ & $m/m_\rho$ & $(m/m_\rho)^{\rm exp.}$ & $m/m_\rho$ \cite{Bali:2013kia} \\\hline
$0^{-+}$ ($\pi$)& 0 & 0 & 0.174|0.180 & 0 \\
$1^{--}$ ($\rho$)& 0.669314 & 1 &1 & 1\\
$1^{++}$ ($a_1$)& 1.568766 & 1.531 & 1.59(5) & 1.86(2) \\
$1^{--}$ ($\rho^*$)& 2.874323 & 2.072 & 1.89(3) & 2.40(4) \\
$1^{++}$ ($a_1^*$)& 4.546104 & 2.606 & 2.12(3) & 2.98(5) \\
\hline
\end{tabular}
% Or use
%\vspace*{5cm}  % with the correct table height
\end{table}

\section{Glueball spectrum and glueball decay rates}

With the above successes as motivation, we now come back to the glueball spectrum
as obtained already in 2000 in Ref.~\cite{Brower:2000rp} and displayed in Fig.~\ref{figGB}a.
Here the scale is determined by $\MKK$ alone. With $\MKK=949$~MeV, the lowest
scalar mode (lowest green line in Fig.~\ref{figGB}a) turns out ot have a mass of $M_G\approx 855$~MeV, only marginally heavier than the $\rho$ meson.
Lattice simulations in QCD give values around 1.7~GeV, a factor of 2 higher.
This mismatch also persists when one compares with the large-$N_c$ lattice results of \cite{Lucini:2010nv}:
$M_G/\sqrt{\sigma}$ is $\approx 1.44$ with the parameters of the Sakai-Sugimoto model, while
Ref.~\cite{Lucini:2010nv} finds $\approx 2.73$ for the lowest scalar glueball (see Fig.~\ref{figGB}b,
where the dotted lines display the holographic results for the glueball spectrum normalized
to $\sqrt{\sigma}$).
The generalization
of the Sakai-Sugimoto model, where the D8 branes are not maximally separated, does not help
in this respect. On the contrary, it increases $m_\rho/\MKK$ while keeping $M_G/\MKK$
fixed. (See Table \ref{tabparams}: For the choice of parameters of Ref.~\cite{Callebaut:2011ab}
for this generalized Sakai-Sugimoto model, $M_G$ is already below $m_\rho$.)

As already discussed, while the glueball spectrum in the Witten model resembles
the spectrum as found on the lattice, it also has some notable qualitative differences
such as a certain proliferation of scalar glueballs $0^{++}$. Moreover, the lowest holographic
glueball comes from an ``exotic'' $g_{44}$ polarization of the metric fluctuation
(while also involving other metric components as well as the dilaton).
The next higher scalar glueball, which has the same mass as the lowest tensor glueball,
does not involve a $g_{44}$ fluctuation, and could thus be viewed as essentially a dilaton fluctuation,
which in simpler bottom-up models \cite{BoschiFilho:2002ta,Colangelo:2007pt,Forkel:2007ru}
is the only way to model a scalar glueball.
The mass of the latter is $M_D=M_T\approx 1.567\MKK\approx 1487$~MeV, which matches reasonably
well with the lowest scalar glueball on the lattice (albeit not for the tensor glueball which
on the lattice is around 2.4--2.6 GeV).

An important open question, which is difficult to address by lattice simulations, is the decay width of
glueballs. In the Witten model, holographic glueballs are stable modes, and they remain so when
one introduces chiral quarks through probe D8 branes. However, as pointed out by Ref.~\cite{Hashimoto:2007ze},
the metric fluctuations dual to glueballs couple to the modes on the D8 brane representing
pions and vector mesons. Ref.~\cite{Hashimoto:2007ze} has considered the decay rate of the lowest
(``exotic'') scalar mode, which was revisited and extended recently by Br\"unner, Parganlija and the
present author \cite{Brunner:2014lya,BPR}.

In order to see whether the Sakai-Sugimoto model has any chance to predict glueball decay rates,
let us begin with considering the decay of the $\rho$ meson first. 
%it is instructive to consider the decay of the $\rho$ meson first. 
The flavor gauge field modes on the D8 brane are all stable,
but the D8-brane action determines all the couplings between different mesons
without further free parameters. To leading order they have been worked out completely in
Ref.~\cite{Sakai:2005yt}, but only a few (particularly interesting and challenging) decay processes have been discussed
in the literature.

It is instructive to consider one of the simplest cases, the decay of the $\rho$ meson, to see whether
one may hope for quantitatively significant predictions. This was done already in Ref.~\cite{Hashimoto:2007ze},
but unfortunately using parameters affected by an error in the normalization of the D8-brane action
present in the original and published version of Ref.~\cite{Sakai:2004cn,Sakai:2005yt} for the multi-flavor case $N_f>1$
(corrected however in the final e-print version). % of Ref.~\cite{Sakai:2004cn,Sakai:2005yt}).
Identifying the mode $n=1$ in (\ref{psin}) with the $\rho$ meson, $A_\mu^{(1)}=\psi_1(Z) v^{(1)}_\mu(x)\equiv
\psi_1(Z) \rho_\mu(x)$ 
and ensuring %normalizing $\psi_1$ such 
that $\rho(x)$ is a canonically normalized vector field yields
\be %c_6 in HTT
\mathcal L_{\rho\pi\pi}=-g_{\rho\pi\pi} %\sqrt2 c_{\rho\pi\pi} 
\epsilon_{abc}(\partial_\mu \pi^a)
\rho^{b\mu}\pi^c,\quad
%c_\rho=
g_{\rho\pi\pi}=\sqrt2
\int dZ \frac1{\pi K}\psi_1=\sqrt{2}\times 24.030%2 
\,\lSS^{-\frac12} N_c^{-\frac12}.
\ee
% where $\lSS\equiv \gYM^2N_c\equiv g^2N_c/2$.
% This agrees with the numerical value given in table 3$\cdot$34 of 
% Ref.~\cite{Sakai:2005yt}
% for $g_{v^1\pi\pi}\equiv g_{\rho\pi\pi}$. %\equiv\sqrt{2}\,c_\rho$.
($g_{\rho\pi\pi}/\sqrt2$ 
was denoted as $c_6$ in \cite{Hashimoto:2007ze}.)

Using the original set of parameters (\ref{kappaMSS}) gives a decay rate of the $\rho$ meson
in two (massless) pions
\be\label{Grhopipi}
\Gamma_\rho/m_\rho=%\frac{c_\rho^2}{24\pi}
\frac{g_{\rho\pi\pi}^2}{48\pi}\approx 7.659\, (\lSS N_c)^{-1}
\approx 0.1535
\ee
which is 20\% lower than the experimental
value $\Gamma_\rho/m_\rho= 0.191(1)$ from the PDG \cite{Agashe:2014kda}.
% (With the incorrectly matched $\lambda$ of  \cite{Hashimoto:2007ze}
% a value of $0.307$ is obtained which is instead larger
% and significantly further off.)
As we have discussed above, the coupling $\lSS\approx 16.63$ appears to be
somewhat too large to fit the QCD string tension with the leading-order result (\ref{sigmastring})
and also the large-$N_c$ lattice result Ref.~\cite{Bali:2013kia} for $m_\rho/\sqrt{\sigma}$.
Using the latter as input 
corresponds to $\lSS\approx 12.55(80)$ and gives an almost perfect prediction of $\Gamma_\rho/m_\rho=0.203(14)$
(if one ignores that $g_{\rho\pi\pi}$ should actually be even a bit larger because
(\ref{Grhopipi}) refers to massless pions).
As suggested in the previous section, we shall also consider the lower value of $12.55$ as an alternative choice of $\lSS$ in order to set up a theoretical error bar.

Turning now to glueball decay rates, one can proceed in analogy to the calculation of the $\rho$ meson decay into
pions \cite{Hashimoto:2007ze,Brunner:2014lya,BPR}. A metric fluctuation $\delta g_{mn}(x^\mu,Z)=G(x)H(Z)\epsilon_{mn}$ dual to a glueball mode needs to be
normalized such that the corresponding scalar, vector, or tensor field $G(x)$ is canonically normalized. 
This fixes the amplitude of the mode function $H$ such that $H(Z=0)^{-1}\propto \lSS^{1/2}N_c\MKK$.
Inserting the metric fluctuations in the D-brane action yields an effective action
for the couplings of glueballs to pions and the vector mesons.
The resulting general pattern is \cite{Hashimoto:2007ze}
\be
\Gamma_{G\to2\pi}/M_G\propto \lSS^{-1}N_c^{-2},\quad
\Gamma_{G\to4\pi}/M_G\propto \lSS^{-3}N_c^{-4},
\ee
so that glueballs are predicted as rather narrow states, with the branching ratio
into 4 pions being particularly strongly suppressed. A decay into four $\pi^0$ is even further
reduced \cite{BPR}:
$F^4$ terms in the DBI action of the D8 branes give vertices of a glueball with four $\pi^0$ such that
\be
\Gamma_{G\to4\pi^0}/M_G\propto \lSS^{-7}N_c^{-4},
\ee
while nonlinear terms of $G$ in the Yang-Mills part of the DBI action give rise to
\be
\Gamma_{G\to G+2\pi^0\to4\pi^0}/M_G\propto \lSS^{-3}N_c^{-6},
\ee
the latter with strong kinematical suppression from phase space integrals.

In Ref.~\cite{Hashimoto:2007ze}, the decay rates of the glueball corresponding to the lowest scalar mode
has been calculated, which, as discussed above, comes out with a mass $M_G\approx 0.9\MKK$,
which is only half of what is expected
from lattice calculations, and which appears exotic in that it involves a metric polarization $g_{44}$.
In Ref.~\cite{Brunner:2014lya,BPR} we have considered also the scalar and tensor glueball associated with
transverse-traceless modes in AdS$_7\times S^4$, which provide glueballs of mass $M_G\approx 1.567\MKK\approx
1487$~MeV and
thus not far from the lowest glueball in lattice simulations. The coupling of this glueball to two pions turns out to read 
\begin{equation}
 S_{G(1487)\to\pi\pi}=\int d^4x\frac{1}{2}\tilde c_1\partial_\mu\pi^a\partial_\nu\pi^a
\left(\eta^{\mu\nu}-\frac{\partial^\mu \partial^\nu}{M_G^2}\right)G,
\end{equation}
with $\tilde c_1=17.23 \lSS^{-1/2} N_c^{-1} M_{\rm KK}^{-1}$, % given by
% \begin{equation}
% \tilde c_1=\frac{17.23}{\lSS^{1/2} N_c M_{\rm KK}}.
% \end{equation}
which yields
\begin{equation}
 \frac{\Gamma_{G(1487)\rightarrow\pi\pi}}{M_G}=\frac{3|\tilde c_1|^2 M_G^2}{512\pi}=\frac{1.359}{\lSS N_c^2}
=\left\{ 0.009 \atop 0.012 \right. \;\mbox{for}\; \lSS=
\left\{ 16.63 \atop 12.55 \right..
\end{equation}
This result is significantly smaller than the relative width obtained for the lowest ``exotic'' scalar mode
(which is of the same parametric order but numerically has a larger coupling\footnote{The relative width has been obtained as 0.040 for $\lSS=16.63$ in Ref.~\cite{Hashimoto:2007ze}, but according to
\cite{BPR} the effective action derived in Ref.~\cite{Hashimoto:2007ze} is incomplete, and
the resulting relative width of the lowest ``exotic'' scalar mode is even larger, namely 0.092.}),
suggesting that the scalar glueball with $M=1487$~MeV does not behave like an excited scalar, which
one would expect to have a larger decay width. This adds to the suspicion that the lowest ``exotic'' scalar
might have to be discarded.

The prediction of the Witten-Sakai-Sugimoto model is then a scalar glueball of roughly the mass
found in lattice calculations and with a very narrow width. This should be contrasted with
the findings of Ref.~\cite{Janowski:2014ppa}, where an extended linear sigma model with a dilaton describing a narrow glueball
could only be accommodated with an uncomfortably large gluon condensate. In the Witten-Sakai-Sugimoto model
a narrow glueball seems to be perfectly compatible with a comparatively small gluon condensate (\ref{C4SS}). 

\section{Deconfinement, chiral symmetry restoration, and the effects of strong magnetic fields}

\subsection{Deconfinement transition to black D4 phase and chiral symmetry restoration}

The Witten-Sakai-Sugimoto model can be extended to finite temperature $T$ by compactifying
also imaginary time $i x^0$ with a period $\beta=1/T$. Periodic boundary conditions for bosons and
antiperiodic boundary conditions for fermions give the correct statistics for the thermodynamic
partition function, which breaks supersymmetry spontaneously exactly as the spatial
circle in $x_4$ does. For low temperature (large $\beta$), the gravitational background
is just (\ref{ds2W}) with imaginary time and $i x^0\simeq ix^0+\beta$. 
As the temperature increases,
the thermal circle shrinks and a phase transition happens when the circumference of the thermal circle
becomes equal to that of the $x_4$-circle, because at temperatures higher than that it is energetically favorable
that the Euclidean black hole with its cigar topology in the $x_4$-$u$ subspace exchanged with the cylindrical
topology in the $ix^0$-$u$ subspace \cite{Aharony:2006da}. 
The phase transition temperature is simply given by
\be\label{Tcdec}
\beta=L_4\equiv 2\pi R_4 \Leftrightarrow T_c=\frac1{2\pi}\MKK. % \approx 151\;{\rm MeV}.
\ee
This corresponds to the appearance of an actual black hole (more precisely
a black 4-brane) in the real-time geometry, analogous to a Hawking-Page transition \cite{Witten:1998zw}. 
Since with this swapping of the roles of $x_4$ and $ix^0$, the blackening
function $f(u)$ now appears in $g_{tt}$, with $\uKK$ replaced by $u_T=(4\pi T/3)^2 R^3$, and the string tension (\ref{sigmastring}) suddenly
becomes zero, signalling deconfinement. 

\begin{figure}[h]
\centerline{\includegraphics[width=0.6\textwidth]{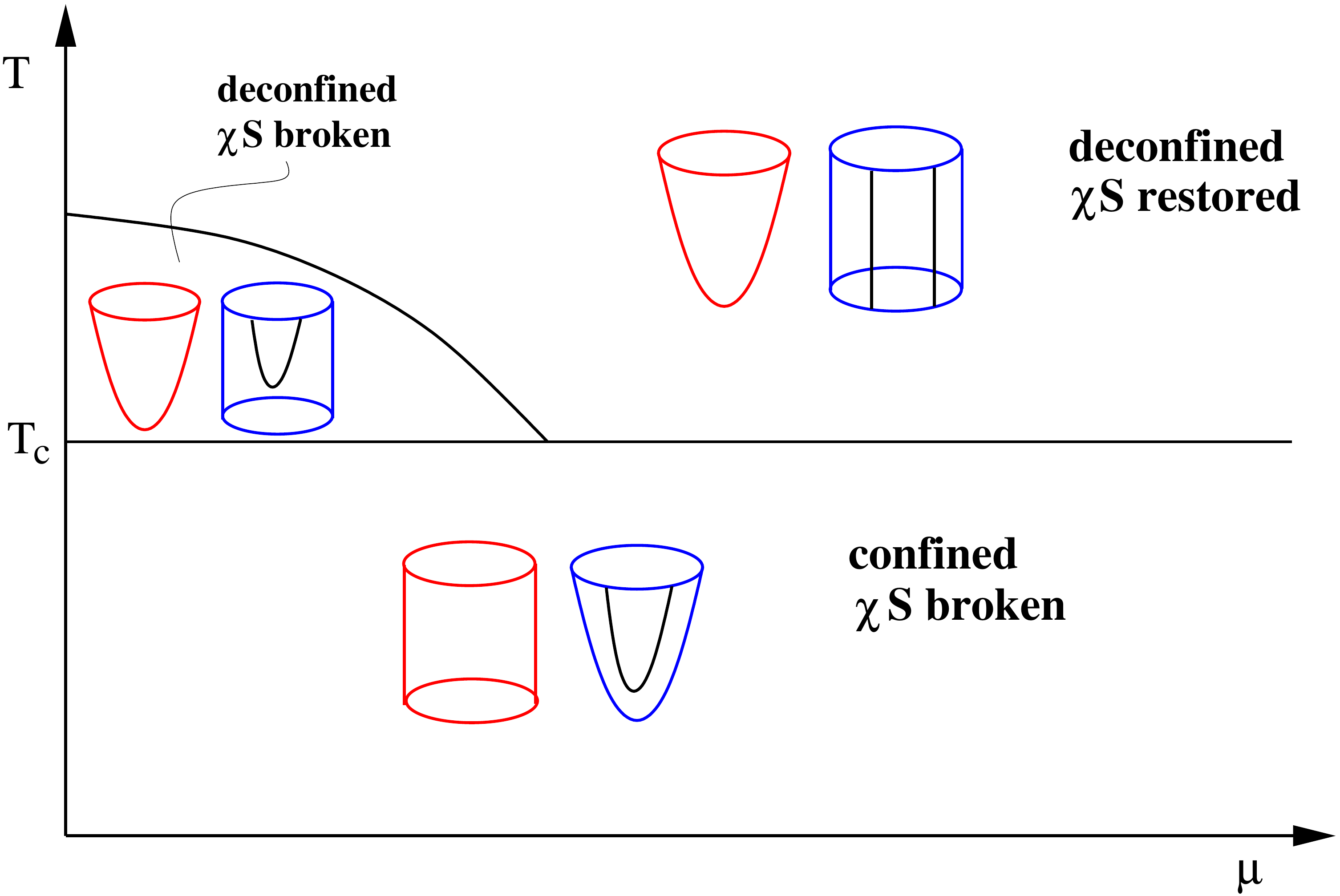}}
\caption{Phase diagram of the Sakai-Sugimoto model where the deconfinement transition is
realized by a Hawking-Page transition. The topology of the $ix^0$-$u$ and $x_4$-$u$ subspaces
is sketched in red and blue, respectively, with the D8-$\overline{\mbox{D8}}$-brane configuration
drawn in black.}
\label{figTmu}       % Give a unique label
\end{figure}

Unfortunately, this deconfinement transition is not smoothly connected to the deconfinement transition
in the 3+1-dimensional Yang-Mills theory and we shall mention an alternative proposal for a more adequate
holographic model below. The deconfined phase corresponding to the black 4-brane is nevertheless interesting, because
it gives rise to a nontrivial pattern of chiral symmetry restoration if one relaxes the original
setup of antipodal D8-$\overline{\mbox{D8}}$-branes \cite{Aharony:2006da}. When the latter are maximally separated,
as in the original Sakai-Sugimoto model, chiral symmetry is necessarily restored when the cigar topology
in the $u$-$x_4$ is replaced by a cylindrical topology. The connection between the D8-$\overline{\mbox{D8}}$-branes
is broken and their configuration becomes that of simple straight embeddings down to $u=u_T$, the black brane horizon.
However, with nonmaximal separation, the D8-$\overline{\mbox{D8}}$-branes can connect also within the cylindrical
topology, and it depends on the free energy encoded in the D8-action which configuration is favored.
It turns out that for brane separation $L<0.97 R_4$ there is still a chiral symmetry breaking phase
when the deconfined phase is entered. If one also introduces a chemical potential for quark number
in the form of an asymptotic value of the U(1)-component of $A^0$, the zeroth component of the flavor
gauge field, the phase diagrams takes the form displayed in Fig.~\ref{figTmu} \cite{Horigome:2006xu}.

This phase diagram has also been studied in the presence of baryonic matter represented by
sources at the tip of the D8 branes \cite{Bergman:2007wp}, and refinements
involving extended, non-point-like baryons have been studied in \cite{Kim:2007vd,Rozali:2007rx,Kaplunovsky:2012gb,Ghoroku:2012am}.
Also a very interesting attempt \cite{deBoer:2012ij} has been made to find a holographic realization of the so-called
quarkyonic phase \cite{McLerran:2007qj,Glozman:2007tv} within the confined phase. % (safe from the problem discussed in the next subsection).

\subsection{Deconfinement through a Gregory-Laflamme transition}

The first-order transition to the black 4-brane phase takes place at a temperature (\ref{Tcdec}) directly set by $\MKK$:
%\be\label{Tcdec}
$T_c=\MKK/2\pi \approx 151\;{\rm MeV}$.
%\ee
Superficially, this looks appealing as the (crossover) temperature for deconfinement and chiral symmetry
breaking of real QCD is indeed close to that value. However, when compared to the first-order phase transition temperature
for pure-glue (or quenched) QCD, it is uncomfortably small.

A more important issue with this deconfinement transition has been noted first in Ref.~\cite{Aharony:2006da}:
In the limit of small 't Hooft coupling and large $\MKK$, which would be required to actually
reach 3+1-dimensional Yang-Mills theory, the critical temperature will have to occur at $T\sim\Lambda_{\rm QCD}\ll\MKK=R_4^{-1}$.
However, due to the symmetry under $\beta\leftrightarrow L_4\equiv 2\pi R_4$, % $T\leftrightarrow 1/2\pi R_4$, 
there will then also be a phase transition at
a temperature $T\gg R_4^{-1}$. The phase transition line given by (\ref{Tcdec}), if it connects to the
deconfinement of 3+1-dimensional Yang-Mills theory, needs to bifurcate at some value of $\lambda_5/R_4$;
it cannot smoothly connect.

In Ref.~\cite{Mandal:2011ws} 
it has recently been pointed out that the $Z_{N_c}\times Z_{N_c}$ center symmetry for the 
two circle compactifications along $ix^0$ and $x_4$ matches the phases of 3+1-dimensional Yang-Mills theory
only in the low-temperature configuration, but not in the high-temperature, black 4-brane geometry.
Hence, there definitely has to be a phase transition between the latter and actual deconfined 3+1-dimensional Yang-Mills theory.
The authors of Ref.~\cite{Mandal:2011ws} have argued that one should instead consider
a phase transition with periodic instead of antiperiodic boundary conditions for fermions on the thermal circle.
This is not appropriate for fermions, but the target of the Witten model is pure Yang-Mills theory without fermions
anyway. In this case, the symmetry $\beta\leftrightarrow L_4$ is avoided
and the gravity dual of the deconfinement
transition is a Gregory-Laflamme transition into the T-dual type IIB supergravity, where the D4 branes are replaced
by an inhomogeneous distribution of D3 branes.
Then the center symmetry breaking is such that it does not rule out a smooth connection to the deconfinement
transition in 3+1-dimensional Yang-Mills theory.

\begin{figure}[h]
\centerline{\includegraphics[width=\textwidth]{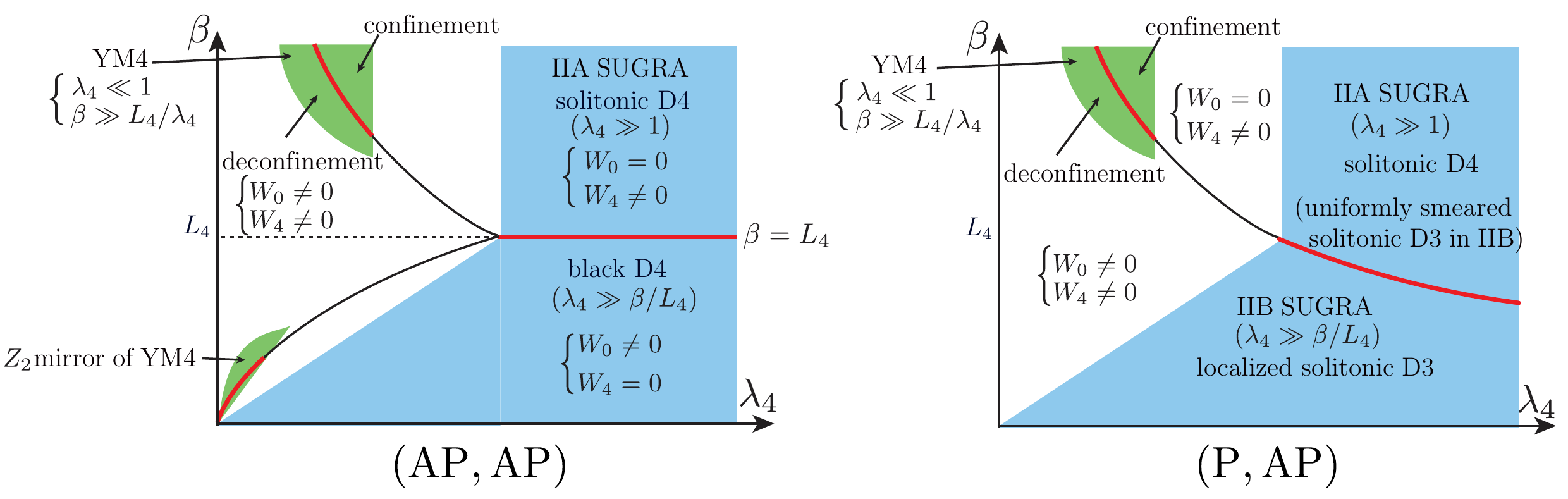}}
\caption{%$L_4=2\pi R_4$. 
Deconfinement phase transitions with antiperiodic and periodic boundary conditions
along the thermal circle, with $W_0$ and $W_4$ referring to the expection values of Polyakov loops
in temporal and $x_4$ direction, respectively (taken from Ref.~\cite{Mandal:2011uq}).}
\label{figD4phase}       % Give a unique label
\end{figure}

Using prior work on the Gregory-Laflamme instability of branes compactified on a circle \cite{Harmark:2004ws},
Ref.~\cite{Mandal:2011ws} has obtained the following estimate for the first-order transition temperature:
\be
T_{\rm GL}\simeq \frac{\lSS}{2\pi\cdot 8.54}\MKK.
\ee
Inserting our fit parameters, this gives
$T_{\rm GL}=294$ (222) MeV for $\lSS=16.63$ (12.55), indeed a quantatively more
appealing ballpark for the deconfinement temperature of quenched QCD than the 
transition temperature (\ref{Tcdec}) with antiperiodic boundary conditions on the thermal circle.
For the fermions that are introduced through probe flavor branes, Ref.~\cite{Mandal:2011ws}
suggested to use an imaginary chemical potential to get the statistics right, but
the details of this alternative proposal for a deconfined Witten-Sakai-Sugimoto model
have still to be worked out.

\subsection{The effects of strong magnetic fields on chiral symmetry breaking and restoration}

While the simple deconfinement transition based on black 4-branes appears to be lacking important
features of the 3+1-dimensional deconfinement of nonsupersymmetric Yang-Mills theory, it may nevertheless
be interesting to study the chiral phase transition in the simple background of black 4-branes.
With nonmaximal separation of the D8 branes, the chiral phase transition is moved away from
the deconfinement transition. For sufficiently small D8 brane separation, the deconfinement temperature
can actually be sent to arbitarily small temperature \cite{Parnachev:2006dn}.
In this limit, the Sakai-Sugimoto model actually becomes dual to a nonlocal NJL model,
as has been pointed out in Ref.~\cite{Antonyan:2006vw}, and 
the phase diagram in the $T$-$\mu$-plane is similar to that found in conventional NJL models.

\begin{figure}[htb]
\centering
\sidecaption
\includegraphics[width=7.35cm,clip]{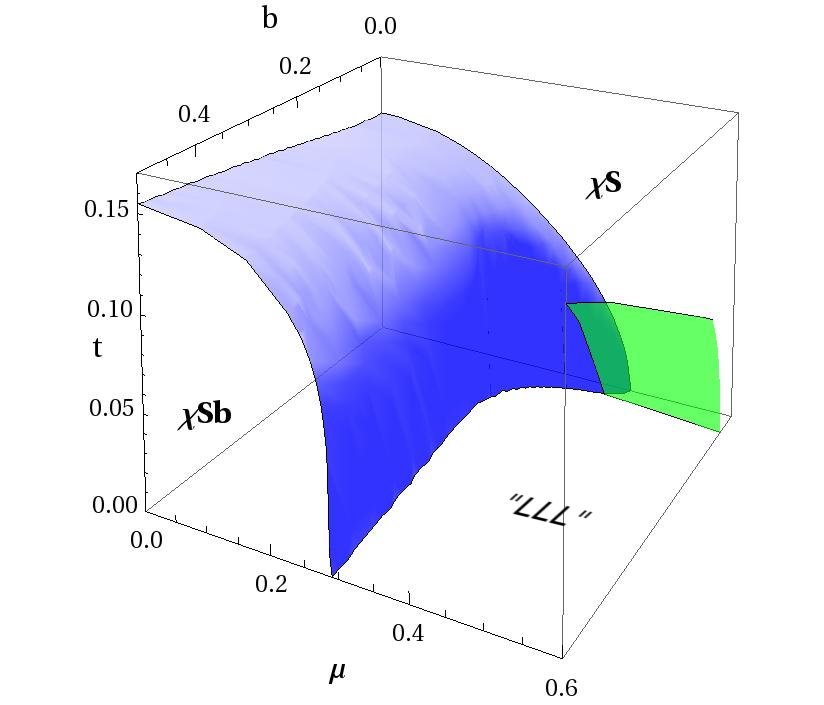}
\caption{Phase diagram of the Sakai-Sugimoto model in (dimensionless) temperature ($t$), quark chemical potential ($\mu$), and magnetic field strength at small D8 brane separation
such that the deconfinement transition is at $t\approx 0$, without baryonic matter
from D4 branes \cite{Preis:2010cq} (the effects of including the latter
have been discussed in \cite{Preis:2011sp}). At small $\mu$, the critical temperature
for chiral symmetry restoration rises with $b$ (normal magnetic catalysis), whereas
at low $T$, the dependence is nonmonotonic. Beyond a critical temperature, there
is also a magnetic phase transition (green surface) into a ``lowest Landau-level'' phase.}       % Give a unique label
\label{fig3D}
\end{figure}

With nonantipodal D8 branes, the phase diagram moreover becomes sensitive to external fields,
which makes this setup an interesting toy model for exploring possible phenomena related to
strong electromagnetic fields, in particular their impact on chiral symmetry breaking.
At zero baryon chemical potential, magnetic catalysis of chiral symmetry breaking
was shown to occur in the Sakai-Sugimoto model \cite{Johnson:2008vna,Bergman:2008sg},
in accordance with the field-theoretic analysis of Ref.~\cite{Gusynin:1994xp}.
At low temperature and high chemical potential, an opposite effect (``inverse
magnetic catalysis'') was observed
in Ref.~\cite{Preis:2010cq,Preis:2012fh} that has to do with an additional
phase transition \cite{Lifschytz:2009sz,Preis:2010cq} that can be interpreted as a transition into the lowest Landau level (LLL)
of chiral quarks (Fig.~\ref{fig3D}).\footnote{At zero chemical potential,
the magnetic field monotonically increases the critical temperature
for chiral symmetry restoration, signalling ordinary magnetic catalysis, in
contrast to the recent lattice results for QCD at physical quark masses \cite{Bali:2011qj}.} Indeed, a very similar phase diagram was previously found
in NJL models \cite{Ebert:1999ht,Inagaki:2003yi}. In contrast to the latter,
the holographic model does not show de Haas-van Alphen oscillations, which indicates that
in the holographic model there is no sharp Fermi surface (see also Refs.~\cite{Kulaxizi:2008jx,DiNunno:2014bxa}).

An interesting feature, shared with real QCD \cite{Son:2007ny}, is that in the chirally broken
phase at finite chemical potential
and with a background magnetic field, baryon number is generated through the axial
anomaly by gradients of the pion field. 
On the holographic side, this is brought about by the Chern-Simons part of
the D8 brane action \cite{Thompson:2008qw,Bergman:2008qv,Rebhan:2008ur}.\footnote{There is
however a still unresolved issue with boundary terms that need to be added to
the action to insure thermodynamic consistency, which comes at the price of
modifying the anomaly content \cite{Bergman:2008qv,Rebhan:2008ur,Rebhan:2009vc}.}

% Magnetic catalysis \cite{Gusynin:1994xp} at zero $\mu$ \cite{Johnson:2008vna,Bergman:2008sg}
% 
% Inverse magnetic catalysis at low $T$ \cite{Preis:2010cq,Preis:2012fh}
% 
% with baryons \cite{Preis:2011sp}
% 
% NJL \cite{Ebert:1999ht,Inagaki:2003yi}

%For figure with sidecaption legend use syntax of figure~\ref{fig-2}.

Another consequence of the axial anomaly is the appearance of anomalous nondissipative
transport phenomena. In the Sakai-Sugimoto model at finite chemical potential and nonzero magnetic field, the chirally
symmetric phase carries an axial current $\mathbf j_A \propto \mu \mathbf B$ in accordance with the 
field-theoretic result obtained in Ref.~\cite{Metlitski:2005pr,Newman:2005as}.

A related effect is the so-called
chiral magnetic effect \cite{Fukushima:2008xe,Kharzeev:2013ffa} that has attracted a lot of attention because
it may be studied in heavy-ion collisions where sufficiently strong magnetic fields
could arise in noncentral collisions. Here an imbalance between left and right chiral fermions,
described by an axial chemical potential, gives rise to a vector current and electric charge
separation through $\mathbf j_V \propto \mu_A \mathbf B$ as first noted in a different context in
Ref.~\cite{Vilenkin:1980fu} and independently rediscovered by various authors.
In the Sakai-Sugimoto model this effect was reproduced in Ref.~\cite{Yee:2009vw}, including
a calculation of the frequency dependent chiral conductivity, but
it involves some conceptual issues \cite{Rebhan:2009vc}
that were clarified only recently \cite{Gynther:2010ed,Jimenez-Alba:2014iia}.

\section{Conclusion}

The Witten-Sakai-Sugimoto model is certainly the best studied and most developed top-down holographic approach to
a nonsupersymmetric nonconformal Yang-Mills theory at strong coupling. As we have discussed, this model
reproduces many phenomena that are known to characterize low-energy QCD such as confinement, chiral symmetry breaking,
mesons including glueballs, baryons, and effects from the axial anomaly. Although this model
is actually a 4+1-dimensional super-Yang-Mills theory above a Kaluza-Klein scale that cannot
be made much larger than 1 GeV without leaving the supergravity approximation, % which makes concrete calculations
%possible in the first place, 
extending the results of these calculations to finite
$N_c=3$ and a finite 't Hooft coupling does give a surprisingly good fit of most of the quantities that
can be compared with experiment or controllable lattice calculations. %Since it hardly has any arbitrary parameters,
%it is very predictive (and falsifiable). 
Parametrically, the predictions are in line with the expectations
from large-$N_c$ analysis, but, optimistically, they also give a rough estimate of the concrete magnitude of various effects.
For example, glueball decay rates are found to be numerically very small, which is an interesting prediction given
that the decay rates of known mesons are quite well reproduced by this model.
This may motivate the inclusion of this and related predictions in other, phenomenological models, if only
as a ballpark estimate. Similarly, the model may serve to give thought-provoking suggestions for the phase diagram
of QCD in regions which will hopefully be explored by first-principle lattice simulations. Indeed,
it seems that the notorious sign problem of lattice QCD at finite chemical potential could be overcome
in the future by new ingenious techniques \cite{Aarts:2014bwa}.

On the other hand, the Witten-Sakai-Sugimoto model, being also a first-principle approach, will hopefully also
see further developments. Important and ambitious directions would be the inclusion of backreaction of flavor branes,
corresponding to an unquenching of the quarks \cite{Burrington:2007qd,Erdmenger:2011bw,Bigazzi:2014qsa},
an improved description of the deconfined phase \cite{Mandal:2011uq}, the inclusion of finite quark masses
\cite{Bergman:2007pm,Dhar:2007bz,Hashimoto:2008sr}, or the inclusion of string-theoretic corrections.

\begin{acknowledgement}
 I would like to thank Frederic Br\"unner, Denis Parganlija, Florian Preis, Andreas Schmitt, and Stefan Stricker for enjoyable collaborations on the Witten-Sakai-Sugimoto model as well as
Matthias Lutz, Marco Panero and Shigeki Sugimoto for discussions.
I am particularly grateful to Frederic Br\"unner, Denis Parganlija, and Florian Preis for a careful reading of this manuscript,
and to Andreas Schmitt for some of the illustrations.
\end{acknowledgement}

%
% BibTeX or Biber users please use (the style is already called in the class, ensure that the "woc.bst" style is in your local directory)
\bibliography{REBHAN_Anton_ICNFP2014}

\begin{thebibliography}{100}

\bibitem{Lutz:2009ff}
M.~Lutz et~al. (PANDA Collaboration) (2009), \texttt{0903.3905}

\bibitem{Wiedner:2011mf}
U.~Wiedner, Prog.Part.Nucl.Phys. \textbf{66}, 477 (2011), \texttt{1104.3961}

\bibitem{Friman:2011zz}
B.~Friman, C.~Hohne, J.~Knoll, S.~Leupold, J.~Randrup et~al., Lect.Notes Phys.
  \textbf{814}, 1 (2011)

\bibitem{Aharony:1999ti}
O.~Aharony, S.S. Gubser, J.M. Maldacena, H.~Ooguri, Y.~Oz, Phys.Rept.
  \textbf{323}, 183 (2000), \texttt{hep-th/9905111}

\bibitem{Witten:1998zw}
E.~Witten, Adv.Theor.Math.Phys. \textbf{2}, 505 (1998), \texttt{hep-th/9803131}

\bibitem{Sakai:2004cn}
T.~Sakai, S.~Sugimoto, Prog.Theor.Phys. \textbf{113}, 843 (2005),
  \texttt{hep-th/0412141}

\bibitem{Sakai:2005yt}
T.~Sakai, S.~Sugimoto, Prog.Theor.Phys. \textbf{114}, 1083 (2005),
  \texttt{hep-th/0507073}

\bibitem{Myers:2008yi}
R.C. Myers, M.F. Paulos, A.~Sinha, Phys.Rev. \textbf{D79}, 041901 (2009),
  \texttt{0806.2156}

\bibitem{Blaizot:2006tk}
J.P. Blaizot, E.~Iancu, U.~Kraemmer, A.~Rebhan, JHEP \textbf{0706}, 035 (2007),
  \texttt{hep-ph/0611393}

\bibitem{Kharzeev:2013jha}
D.~Kharzeev, K.~Landsteiner, A.~Schmitt, H.U. Yee, Lect.Notes Phys.
  \textbf{871}, 1 (2013)

\bibitem{Peeters:2007ab}
K.~Peeters, M.~Zamaklar, Eur.Phys.J.ST \textbf{152}, 113 (2007),
  \texttt{0708.1502}

\bibitem{'tHooft:1973jz}
G.~'t~Hooft, Nucl.Phys. \textbf{B72}, 461 (1974)

\bibitem{Maldacena:1997re}
J.M. Maldacena, Int.J.Theor.Phys. \textbf{38}, 1113 (1999),
  \texttt{hep-th/9711200}

\bibitem{Susskind:1994vu}
L.~Susskind, J.Math.Phys. \textbf{36}, 6377 (1995), \texttt{hep-th/9409089}

\bibitem{Policastro:2001yc}
G.~Policastro, D.T. Son, A.O. Starinets, Phys.Rev.Lett. \textbf{87}, 081601
  (2001), \texttt{hep-th/0104066}

\bibitem{Kovtun:2004de}
P.~Kovtun, D.T. Son, A.O. Starinets, Phys.Rev.Lett. \textbf{94}, 111601 (2005),
  \texttt{hep-th/0405231}

\bibitem{Casalderrey-Solana:2014wca}
J.~Casalderrey-Solana, D.C. Gulhan, J.G. Milhano, D.~Pablos, K.~Rajagopal,
  Nucl.Phys. \textbf{A} (2014), \texttt{1408.5616}

\bibitem{Iancu:2014ava}
E.~Iancu, A.~Mukhopadhyay (2014), \texttt{1410.6448}

\bibitem{Freund:1980xh}
P.G. Freund, M.A. Rubin, Phys.Lett. \textbf{B97}, 233 (1980)

\bibitem{Kanitscheider:2008kd}
I.~Kanitscheider, K.~Skenderis, M.~Taylor, JHEP \textbf{0809}, 094 (2008),
  \texttt{0807.3324}

\bibitem{Brower:2000rp}
R.C. Brower, S.D. Mathur, C.I. Tan, Nucl.Phys. \textbf{B587}, 249 (2000),
  \texttt{hep-th/0003115}

\bibitem{Lucini:2010nv}
B.~Lucini, A.~Rago, E.~Rinaldi, JHEP \textbf{1008}, 119 (2010),
  \texttt{1007.3879}

\bibitem{Csaki:1998qr}
C.~Csaki, H.~Ooguri, Y.~Oz, J.~Terning, JHEP \textbf{9901}, 017 (1999),
  \texttt{hep-th/9806021}

\bibitem{Csaki:1999vb}
C.~Csaki, J.~Russo, K.~Sfetsos, J.~Terning, Phys.Rev. \textbf{D60}, 044001
  (1999), \texttt{hep-th/9902067}

\bibitem{Constable:1999gb}
N.R. Constable, R.C. Myers, JHEP \textbf{9910}, 037 (1999),
  \texttt{hep-th/9908175}

\bibitem{Kruczenski:2003uq}
M.~Kruczenski, D.~Mateos, R.C. Myers, D.J. Winters, JHEP \textbf{0405}, 041
  (2004), \texttt{hep-th/0311270}

\bibitem{Burrington:2007qd}
B.A. Burrington, V.S. Kaplunovsky, J.~Sonnenschein, JHEP \textbf{0802}, 001
  (2008), \texttt{0708.1234}

\bibitem{Bigazzi:2014qsa}
F.~Bigazzi, A.L. Cotrone (2014), \texttt{1410.2443}

\bibitem{Barbon:2004dq}
J.L. Barbon, C.~Hoyos-Badajoz, D.~Mateos, R.C. Myers, JHEP \textbf{0410}, 029
  (2004), \texttt{hep-th/0404260}

\bibitem{Bali:2013kia}
G.S. Bali, F.~Bursa, L.~Castagnini, S.~Collins, L.~Del~Debbio et~al., JHEP
  \textbf{1306}, 071 (2013), \texttt{1304.4437}

\bibitem{Teper:1997am}
M.J. Teper (1997), \texttt{hep-lat/9711011}

\bibitem{Shifman:1978bx}
M.A. Shifman, A.~Vainshtein, V.I. Zakharov, Nucl.Phys. \textbf{B147}, 385
  (1979)

\bibitem{Ioffe:2005ym}
B.~Ioffe, Prog.Part.Nucl.Phys. \textbf{56}, 232 (2006), \texttt{hep-ph/0502148}

\bibitem{Samsonov:2004zm}
A.~Samsonov (2004), \texttt{hep-ph/0407199}

\bibitem{Narison:2011xe}
S.~Narison, Phys.Lett. \textbf{B706}, 412 (2012), \texttt{1105.2922}

\bibitem{Bali:2014sja}
G.S. Bali, C.~Bauer, A.~Pineda, Phys.Rev.Lett. \textbf{113}, 092001 (2014),
  \texttt{1403.6477}

\bibitem{Colangelo:2010et}
G.~Colangelo, S.~Durr, A.~Juttner, L.~Lellouch, H.~Leutwyler et~al.,
  Eur.Phys.J. \textbf{C71}, 1695 (2011), \texttt{1011.4408}

\bibitem{Imoto:2010ef}
T.~Imoto, T.~Sakai, S.~Sugimoto, Prog.Theor.Phys. \textbf{124}, 263 (2010),
  \texttt{1005.0655}

\bibitem{Schechter:1999hg}
J.~Schechter, H.~Weigel (1999), \texttt{hep-ph/9907554}

\bibitem{Witten:1998xy}
E.~Witten, JHEP \textbf{9807}, 006 (1998), \texttt{hep-th/9805112}

\bibitem{Son:2003et}
D.~Son, M.~Stephanov, Phys.Rev. \textbf{D69}, 065020 (2004),
  \texttt{hep-ph/0304182}

\bibitem{Hong:2007kx}
D.K. Hong, M.~Rho, H.U. Yee, P.~Yi, Phys.Rev. \textbf{D76}, 061901 (2007),
  \texttt{hep-th/0701276}

\bibitem{Hata:2007mb}
H.~Hata, T.~Sakai, S.~Sugimoto, S.~Yamato, Prog.Theor.Phys. \textbf{117}, 1157
  (2007), \texttt{hep-th/0701280}

\bibitem{Hashimoto:2008zw}
K.~Hashimoto, T.~Sakai, S.~Sugimoto, Prog.Theor.Phys. \textbf{120}, 1093
  (2008), \texttt{0806.3122}

\bibitem{Seki:2008mu}
S.~Seki, J.~Sonnenschein, JHEP \textbf{0901}, 053 (2009), \texttt{0810.1633}

\bibitem{Adkins:1983ya}
G.S. Adkins, C.R. Nappi, E.~Witten, Nucl.Phys. \textbf{B228}, 552 (1983)

\bibitem{Antonyan:2006vw}
E.~Antonyan, J.~Harvey, S.~Jensen, D.~Kutasov (2006), \texttt{hep-th/0604017}

\bibitem{Aharony:2006da}
O.~Aharony, J.~Sonnenschein, S.~Yankielowicz, Annals Phys. \textbf{322}, 1420
  (2007), \texttt{hep-th/0604161}

\bibitem{Callebaut:2011ab}
N.~Callebaut, D.~Dudal, H.~Verschelde, JHEP \textbf{1303}, 033 (2013),
  \texttt{1105.2217}

\bibitem{BoschiFilho:2002ta}
H.~Boschi-Filho, N.R. Braga, Eur.Phys.J. \textbf{C32}, 529 (2004),
  \texttt{hep-th/0209080}

\bibitem{Colangelo:2007pt}
P.~Colangelo, F.~De~Fazio, F.~Jugeau, S.~Nicotri, Phys.Lett. \textbf{B652}, 73
  (2007), \texttt{hep-ph/0703316}

\bibitem{Forkel:2007ru}
H.~Forkel, Phys.Rev. \textbf{D78}, 025001 (2008), \texttt{0711.1179}

\bibitem{Hashimoto:2007ze}
K.~Hashimoto, C.I. Tan, S.~Terashima, Phys.Rev. \textbf{D77}, 086001 (2008),
  \texttt{0709.2208}

\bibitem{Brunner:2014lya}
F.~Brünner, D.~Parganlija, A.~Rebhan, Acta Phys.Polon.Supp. \textbf{7}, 533
  (2014), \texttt{1407.6914}

\bibitem{BPR}
F.~Brünner, D.~Parganlija, A.~Rebhan (2014), in preparation

\bibitem{Agashe:2014kda}
K.~Olive et~al. (Particle Data Group), Chin.Phys. \textbf{C38}, 090001 (2014)

\bibitem{Janowski:2014ppa}
S.~Janowski, F.~Giacosa, D.H. Rischke (2014), \texttt{1408.4921}

\bibitem{Horigome:2006xu}
N.~Horigome, Y.~Tanii, JHEP \textbf{0701}, 072 (2007), \texttt{hep-th/0608198}

\bibitem{Bergman:2007wp}
O.~Bergman, G.~Lifschytz, M.~Lippert, JHEP \textbf{0711}, 056 (2007),
  \texttt{0708.0326}

\bibitem{Kim:2007vd}
K.Y. Kim, S.J. Sin, I.~Zahed, JHEP \textbf{0809}, 001 (2008),
  \texttt{0712.1582}

\bibitem{Rozali:2007rx}
M.~Rozali, H.H. Shieh, M.~Van~Raamsdonk, J.~Wu, JHEP \textbf{0801}, 053 (2008),
  \texttt{0708.1322}

\bibitem{Kaplunovsky:2012gb}
V.~Kaplunovsky, D.~Melnikov, J.~Sonnenschein, JHEP \textbf{1211}, 047 (2012),
  \texttt{1201.1331}

\bibitem{Ghoroku:2012am}
K.~Ghoroku, K.~Kubo, M.~Tachibana, T.~Taminato, F.~Toyoda, Phys.Rev.
  \textbf{D87}, 066006 (2013), \texttt{1211.2499}

\bibitem{deBoer:2012ij}
J.~de~Boer, B.D. Chowdhury, M.P. Heller, J.~Jankowski, Phys.Rev. \textbf{D87},
  066009 (2013), \texttt{1209.5915}

\bibitem{McLerran:2007qj}
L.~McLerran, R.D. Pisarski, Nucl.Phys. \textbf{A796}, 83 (2007),
  \texttt{0706.2191}

\bibitem{Glozman:2007tv}
L.Y. Glozman, R.~Wagenbrunn, Phys.Rev. \textbf{D77}, 054027 (2008),
  \texttt{0709.3080}

\bibitem{Mandal:2011ws}
G.~Mandal, T.~Morita, JHEP \textbf{1109}, 073 (2011), \texttt{1107.4048}

\bibitem{Mandal:2011uq}
G.~Mandal, T.~Morita, J.Phys.Conf.Ser. \textbf{343}, 012079 (2012),
  \texttt{1111.5190}

\bibitem{Harmark:2004ws}
T.~Harmark, N.A. Obers, JHEP \textbf{0409}, 022 (2004), \texttt{hep-th/0407094}

\bibitem{Parnachev:2006dn}
A.~Parnachev, D.A. Sahakyan, Phys.Rev.Lett. \textbf{97}, 111601 (2006),
  \texttt{hep-th/0604173}

\bibitem{Johnson:2008vna}
C.V. Johnson, A.~Kundu, JHEP \textbf{0812}, 053 (2008), \texttt{0803.0038}

\bibitem{Bergman:2008sg}
O.~Bergman, G.~Lifschytz, M.~Lippert, JHEP \textbf{0805}, 007 (2008),
  \texttt{0802.3720}

\bibitem{Gusynin:1994xp}
V.~Gusynin, V.~Miransky, I.~Shovkovy, Phys.Lett. \textbf{B349}, 477 (1995),
  \texttt{hep-ph/9412257}

\bibitem{Preis:2010cq}
F.~Preis, A.~Rebhan, A.~Schmitt, JHEP \textbf{1103}, 033 (2011),
  \texttt{1012.4785}

\bibitem{Preis:2012fh}
F.~Preis, A.~Rebhan, A.~Schmitt, Lect.Notes Phys. \textbf{871}, 51 (2013),
  \texttt{1208.0536}

\bibitem{Lifschytz:2009sz}
G.~Lifschytz, M.~Lippert, Phys.Rev. \textbf{D80}, 066007 (2009),
  \texttt{0906.3892}

\bibitem{Bali:2011qj}
G.~Bali, F.~Bruckmann, G.~Endrodi, Z.~Fodor, S.~Katz et~al., JHEP
  \textbf{1202}, 044 (2012), \texttt{1111.4956}

\bibitem{Ebert:1999ht}
D.~Ebert, K.~Klimenko, M.~Vdovichenko, A.~Vshivtsev, Phys.Rev. \textbf{D61},
  025005 (2000), \texttt{hep-ph/9905253}

\bibitem{Inagaki:2003yi}
T.~Inagaki, D.~Kimura, T.~Murata, Prog.Theor.Phys. \textbf{111}, 371 (2004),
  \texttt{hep-ph/0312005}

\bibitem{Kulaxizi:2008jx}
M.~Kulaxizi, A.~Parnachev, Nucl.Phys. \textbf{B815}, 125 (2009),
  \texttt{0811.2262}

\bibitem{DiNunno:2014bxa}
B.S. DiNunno, M.~Ihl, N.~Jokela, J.F. Pedraza, JHEP \textbf{1404}, 149 (2014),
  \texttt{1403.1827}

\bibitem{Son:2007ny}
D.~Son, M.~Stephanov, Phys.Rev. \textbf{D77}, 014021 (2008), \texttt{0710.1084}

\bibitem{Thompson:2008qw}
E.G. Thompson, D.T. Son, Phys.Rev. \textbf{D78}, 066007 (2008),
  \texttt{0806.0367}

\bibitem{Bergman:2008qv}
O.~Bergman, G.~Lifschytz, M.~Lippert, Phys.Rev. \textbf{D79}, 105024 (2009),
  \texttt{0806.0366}

\bibitem{Rebhan:2008ur}
A.~Rebhan, A.~Schmitt, S.A. Stricker, JHEP \textbf{0905}, 084 (2009),
  \texttt{0811.3533}

\bibitem{Rebhan:2009vc}
A.~Rebhan, A.~Schmitt, S.A. Stricker, JHEP \textbf{1001}, 026 (2010),
  \texttt{0909.4782}

\bibitem{Preis:2011sp}
F.~Preis, A.~Rebhan, A.~Schmitt, J.Phys. \textbf{G39}, 054006 (2012),
  \texttt{1109.6904}

\bibitem{Metlitski:2005pr}
M.A. Metlitski, A.R. Zhitnitsky, Phys.Rev. \textbf{D72}, 045011 (2005),
  \texttt{hep-ph/0505072}

\bibitem{Newman:2005as}
G.~Newman, D.~Son, Phys.Rev. \textbf{D73}, 045006 (2006),
  \texttt{hep-ph/0510049}

\bibitem{Fukushima:2008xe}
K.~Fukushima, D.E. Kharzeev, H.J. Warringa, Phys.Rev. \textbf{D78}, 074033
  (2008), \texttt{0808.3382}

\bibitem{Kharzeev:2013ffa}
D.E. Kharzeev, Prog.Part.Nucl.Phys. \textbf{75}, 133 (2014), \texttt{1312.3348}

\bibitem{Vilenkin:1980fu}
A.~Vilenkin, Phys.Rev. \textbf{D22}, 3080 (1980)

\bibitem{Yee:2009vw}
H.U. Yee, JHEP \textbf{0911}, 085 (2009), \texttt{0908.4189}

\bibitem{Gynther:2010ed}
A.~Gynther, K.~Landsteiner, F.~Pena-Benitez, A.~Rebhan, JHEP \textbf{1102}, 110
  (2011), \texttt{1005.2587}

\bibitem{Jimenez-Alba:2014iia}
A.~Jimenez-Alba, K.~Landsteiner, L.~Melgar (2014), \texttt{1407.8162}

\bibitem{Aarts:2014bwa}
G.~Aarts, E.~Seiler, D.~Sexty, I.O. Stamatescu (2014), \texttt{1408.3770}

\bibitem{Erdmenger:2011bw}
J.~Erdmenger, V.G. Filev, D.~Zoakos, JHEP \textbf{1208}, 004 (2012),
  \texttt{1112.4807}

\bibitem{Bergman:2007pm}
O.~Bergman, S.~Seki, J.~Sonnenschein, JHEP \textbf{0712}, 037 (2007),
  \texttt{0708.2839}

\bibitem{Dhar:2007bz}
A.~Dhar, P.~Nag, JHEP \textbf{0801}, 055 (2008), \texttt{0708.3233}

\bibitem{Hashimoto:2008sr}
K.~Hashimoto, T.~Hirayama, F.L. Lin, H.U. Yee, JHEP \textbf{0807}, 089 (2008),
  \texttt{0803.4192}

\end{thebibliography}
% \begin{thebibliography}{}
% %
% % and use \bibitem to create references.
% % Format for Journal Reference
% Journal Author, Journal \textbf{Volume}, page numbers (year)
% Book Author, \textit{Book title} (Publisher, place, year) page numbers
% \end{thebibliography}

\end{document}